\documentclass[
 reprint,
 amsmath,amssymb,
 aps,prl
]{revtex4-2}

\usepackage[dvipsnames, svgnames, x11names]{xcolor} 
\usepackage[T1]{fontenc}
\usepackage{extarrows}
\usepackage{diagbox}
\usepackage{cancel}
\usepackage{multirow}
\usepackage{physics}
\usepackage{setspace}
\usepackage[colorlinks,anchorcolor=blue,linkcolor=blue,citecolor=blue]{hyperref}
\usepackage{mathrsfs}
\usepackage{amsmath}
\usepackage{graphicx}
\usepackage{dcolumn}
\usepackage{bm}
\usepackage{amsmath}

\graphicspath{{figures/}}

\begin{document}

\allowdisplaybreaks[4]
\preprint{APS/123-QED}

\title{\bfseries Most Subradiant Bound Photon Pairs from Chirality-Mediated Dispersion Softening}

\author{Kailin Tan$^{1,2}$}
 \email{kltan22@m.fudan.edu.cn}

\author{Xuanbing Jiang$^{1,2}$}

\author{Dong Wang$^{1,2}$}

\author{Saijun Wu$^{1,2}$}

\affiliation{$^1$Department of Physics, State Key Laboratory of Surface Physics and Key Laboratory of Micro and Nano Photonic Structures (Ministry of Education), Fudan University, Shanghai 200433, China.\\
$^2$Shanghai Key Laboratory of Metasurfaces for Light Manipulation, Shanghai 200433, China.}

\begin{abstract}
    We study the subradiant bound states (BSs) in a two-level atom array chirally coupled to a one-dimensional waveguide. We demonstrate that the chiral interaction can drive BSs to become the most subradiant two-excitation states across a wide spacing range. This phenomenon is rooted in a mechanism of chirality-mediated dispersion softening, where the BS band distortion suppresses the band curvature $|\alpha_2|$ at an extremum point. We rigorously prove that the BS decay rate follows the scaling $\Gamma \sim |\alpha_2|/N^3$, revealing that the reduction of $|\alpha_2|$ is key to suppressing emission and enhancing subradiance. We also show the existence of chiral BSs in a realistic nanofiber interface.
\end{abstract}
    
\maketitle

\textit{Introduction.\textemdash}Collective emission from subwavelength atom arrays, pioneered by Dicke~\cite{dicke1954coherence}, is a cornerstone of modern quantum optics and a key enabler of engineered photon-photon interactions~\cite{chang2014quantum,chang2018colloquium,gonzalez2024light}. By interfacing atom arrays with photonic reservoirs, one can tailor exotic photon correlations and effective nonlinearities, unlocking applications such as photonic logic gates~\cite{bjorn2022passive,tomas2025passive} and photon blockades~\cite{prasad2020correlating,lu2025chiral}. Among these platforms, atom arrays in free space~\cite{asenjo2017exponential,perczel2017topological,needham2019subradiance,rui2020subradiant,moreno2021quantum,shah2024quantum} and those coupled to nanophotonic waveguides (i.e., waveguide QED)~\cite{chang2012cavity,paulisch2016universal,asenjo2017atom,corzo2019waveguide,dhordjevic2021entanglement,sheremet2023waveguide,tabares2023variational,sunami2025scalable,moore2025van} are particularly intriguing due to the long-range interactions. In the multi-excitation regime, which is intimately connected to optical nonlinearity, these one-dimensional (1D) systems universally host two distinct classes of subradiant eigenstates where the destructive interference stably suppresses the collective decay: free-Fermion states (FSs) and bound photon pairs. FSs emerge from the quadratic band extrema of the single-excitation polaritonic dispersion~\cite{zhang2022free}, displaying a universal $N^{-3}$ decay rate scaling with respect to the atom number $N$~\cite{asenjo2017exponential,albrecht2019subradiant,henriet2019critical,zhang2019theory}. Conversely, bound photon pairs are bound states (BSs) of atomic excitations, acting as the microscopic constituents of self-induced-transparency solitons~\cite{mahmoodian2020dynamics,calajo2022emergence}. Two-excitation BSs reside within the bandgaps of the scattering continuum~\cite{bakkensen2021photonic,schrinski2022polariton,calajo2022emergence,poddubny2026bound}, achieving an $N^{-3}$ or even steeper decay rate scaling in waveguide QED~\cite{zhang2020subradiantbound,poddubny2020quasiflat}. The introduction of chiral interactions~\cite{mitsch2014quantum,pichler2015quantum,corzo2016large,le2017nanofiber,jones2020collectively,liedl2024observation,bach2026emergence} further enriches this landscape. For instance, chiral couplings can localize BSs at the edge opposite to the preferred emission direction, revealing a novel non-Hermitian skin effect~\cite{shi2025chiral}.  Subradiant BSs are also a ubiquitous feature in two-dimensional atom arrays~\cite{marques2021bound,tevcer2024strongly}.

Generally, the multi-excitation order in long-time dynamics of waveguide QED is governed by the highly subradiant states~\cite{zhang2020subradiantbound,poddubny2020quasiflat,zhang2022free}. Within the two-excitation manifold, it remains an open question whether the most subradiant states are BSs or FSs, particularly in the presence of chirality. In non-chiral systems, previous studies suggest that the longest-lived two-excitation eigenstates are typically FSs~\cite{asenjo2017exponential,albrecht2019subradiant,zhang2019theory,henriet2019critical,zhang2020subradiantbound}, while BSs can become more subradiant at specific lattice spacings~\cite{zhang2020subradiantbound,poddubny2020quasiflat}. However, the precise mechanisms regarding this subradiant competition have not been fully clarified. In particular, a rigorous analytical derivation for the $N^{-3}$ scaling of BS decay rates, in parallel with the FS counterpart~\cite{zhang2020subradiant,zhang2022free}, has not yet been established.

\begin{figure}[htbp]
    \includegraphics[scale=0.42]{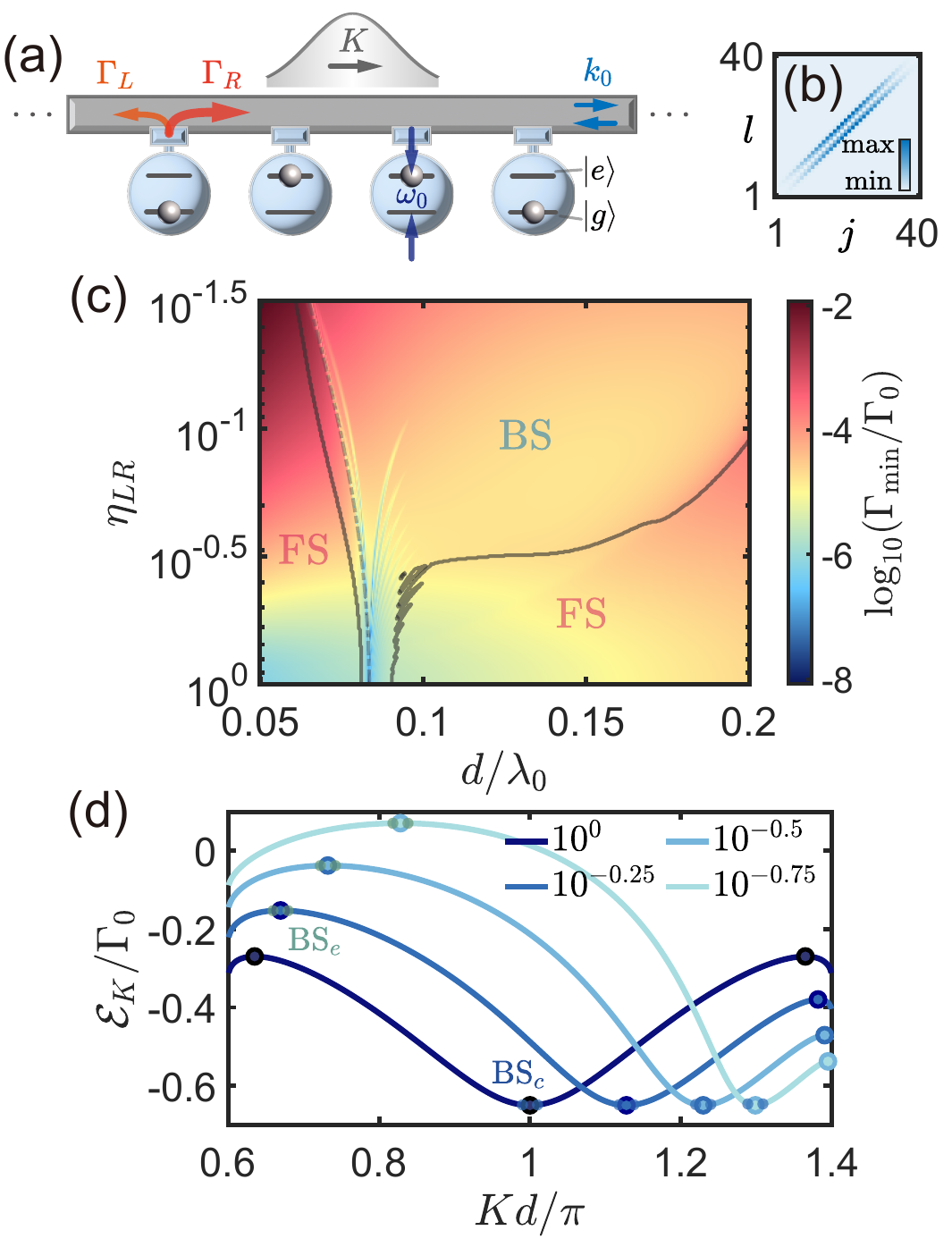}
    \centering
    \caption{(a) Schematic of the chiral waveguide QED model. (b) Probability distribution of a subradiant $\mathrm{BS}_e$ for $N=40$, $d=0.15\lambda_0$ and $\eta_{LR}=10^{-0.75}$. (c) Decay rate distribution of the most subradiant two-excitation state with respect to the lattice spacing $d$ and the asymmetry parameter $\eta_{LR}$ for $N=100$.
    The solid curves separate regions where either the FS or the BS is the most subradiant. The dashed dotted curve corresponds to the emergence of a BS band inflection point with $\alpha_2=0$ (cf. Fig.~\ref{fg:BSinfl}). (d) BS dispersion relation for different values of $\eta_{LR}$ indicated by the legend with $d=0.15\lambda_0$.
    Circles show the energies of $\mathrm{BS}_c$ and $\mathrm{BS}_e$ near the band extrema for an $N=100$ array.} 
    \label{fg:scenario} 
\end{figure}

In this Letter, we demonstrate that chirality can dramatically suppress the emission of BSs, enabling them to surpass FSs to become the most subradiant two-excitation states across a wide range of lattice spacings (see Fig.~\ref{fg:scenario}). To elucidate the underlying physical mechanism, we analytically prove that if the BS dispersion is $\alpha_2(K-K_\mathrm{ex})^2$ near a band extremum $K_\mathrm{ex}$, the corresponding decay rate follows $\Gamma\sim |\alpha_2|/N^3$. The results observed in Fig.~\ref{fg:scenario} is thus driven by a phenomenon we term \textit{dispersion softening}: directional interaction breaks the parity symmetry of the system and distorts the BS band, significantly diminishing the band curvature $|\alpha_2|$ at one of the extrema and strongly reducing the decay rates. We further show the emergence of subradiant chiral BSs in a realistic optical nanofiber interface.

\textit{Subradiant bound states.\textemdash}We consider an array of $N$ two-level atoms chirally coupled to a 1D waveguide, with lattice constant $d$ and guided mode wavevector $k_0$. The non-Hermitian effective Hamiltonian reads~\cite{pichler2015quantum}
\begin{equation}
    \begin{split}
        H_\mathrm{eff}&=-\frac{i}{2}\sum\limits_{j<l}^N e^{ik_0(x_l-x_j)} \left( \Gamma_L \sigma_j^\dagger \sigma_l +  \Gamma_R \sigma_l^\dagger \sigma_j \right)\\
        &\quad -i\frac{\Gamma_0}{2}\sum\limits_{j=1}^N \sigma_j^\dagger \sigma_j.
    \end{split}
    \label{eq:Heff}
\end{equation}
Here, $\Gamma_{L(R)}$ is the decay rate of a single atom into the left- (right-) propagating mode and $\Gamma_0=(\Gamma_L+\Gamma_R)/2$; $\sigma_j^{(\dagger)}$ is the spin lowering (raising) operator of the $j$th atom at $x_j=jd$. We introduce the asymmetry parameter $\eta_{LR}=\Gamma_L/\Gamma_R$ to quantify the chirality. The basis $\ket*{K,\Delta}=\sum_{X/d=1}^{N-\Delta/d} e^{iK(X+\frac{\Delta}{2})} \sigma_X^\dagger\sigma_{X+\Delta}^\dagger\ket*{G}$, characterized by the center-of-excitation momentum $K$ and pair separation $\Delta$, conveniently captures the two-excitation physics, which yields
\begin{equation}
    H_\mathrm{eff}\ket*{K,\Delta}=\sum\limits_{\Delta'/d=1}^{N-1} \mathcal{H}_{\Delta\Delta'}^K \ket*{K,\Delta'}+\ket*{T_{K,\Delta}},
    \label{eq:HeffonKDelta}
\end{equation}
where the effective single-particle Hamiltonian is
\begin{equation}
    \begin{split}
        \mathcal{H}_{\Delta\Delta'}^K = -\frac{i}{2}\sum\limits_{\epsilon=\pm1}&\left[\Gamma_L e^{i\left(k_0+\frac{K}{2}\right)|\Delta+\epsilon\Delta'|}\right.\\
        &\quad \left.
        + \Gamma_R e^{i\left(k_0-\frac{K}{2}\right)|\Delta+\epsilon\Delta'|}\right].
    \end{split}
\end{equation}
The tail $\ket*{T_{K,\Delta}}$, detailed in the Supplemental Material~\cite{SM}, arises from the breakdown of translational invariance due to the finite array length. Following Ref.~\cite{bakkensen2021photonic}, we diagonalize $\mathcal{H}_{\Delta\Delta'}^K$ and find the (normalized) bound eigenstate $\ket*{\psi_K}\sim\sum_{\Delta/d=1}^{N-1}C_\Delta^K\ket*{K,\Delta}$, where $C_\Delta^K=A_K (z_1^K)^{\Delta/d}+B_K (z_2^K)^{\Delta/d}$, with complex roots $z_{1,2}^K$ satisfying $|z_{1,2}^K|<1$. The finite-size correction to $C_\Delta^K$ compared to the infinite-array solution is exponentially small~\cite{SM}. Hence, we obtain
\begin{equation}
    H_\mathrm{eff}\ket*{\psi_K}=\mathcal{E}_K\ket*{\psi_K}+\ket*{T_L}+\ket*{T_R}.
    \label{eq:HeffonpsiK}
\end{equation}

As detailed in the Supplemental Material~\cite{SM}, the tails $\ket*{T_{L,R}}$ are strongly localized at the array edges, representing finite-size scattering effects that prevent $\ket*{\psi_K}$ from being an exact eigenstate of $H_\mathrm{eff}$.

The complexity of $\ket*{T_{L,R}}$ precludes a direct analytical treatment of $H_\mathrm{eff}\ket*{\psi_K}$. In order to proceed, we introduce two key simplifications. First, numerical observations indicate that the most subfradiant BSs are located near the extremum point $K_\mathrm{ex}$. Thus, we focus specifically on the regime where $\delta K=K-K_\mathrm{ex}\sim\mathcal{O}(N^{-1})$ for $N\gg 1$. Note that $\delta K$ is discrete with a separation $\pi/Nd$ in the leading order, due to the finite array length. Second, inspired by Refs.~\cite{zhang2020subradiant,zhang2022free}, we construct a simpler Hamiltonian $\mathbf{H}$ that approximates $H_\mathrm{eff}$ by replacing the slowly varying phase factors of $\ket*{T_{L,R}}$ contributed by $\delta K$ with constant envelopes $e^{i\tilde{\varphi}_{L,R}}=e^{i\delta K Nd \mathbf{f}_{L,R}}$. Here, $\mathrm{Re}\tilde{\varphi}_{L,R}$ represent the average phase shifts, while $\mathrm{Im}\tilde{\varphi}_{L,R}$ ensure population conservation. Specifically, $\mathbf{H}$ has
\begin{equation}
    \mathbf{H}\ket*{\psi_K}=\mathcal{E}_K\ket*{\psi_K}+\mathcal{V}_K\sigma_{k_0}^\dagger\Psi_L^\dagger\ket*{G}+\mathcal{W}_K\sigma_{-k_0}^\dagger\Psi_R^\dagger\ket*{G},
    \label{eq:simplerH}
\end{equation}
with $\mathcal{V}_K=(i\Gamma_R/2)e^{i\tilde{\varphi}_L}$ and $\mathcal{W}_K=(i\Gamma_L/2)e^{i\tilde{\varphi}_R}$. Here, $\sigma_k^\dagger=N^{-1/2} \sum_{j=1}^N e^{ikx_j} \sigma_j^\dagger$ denotes the timed-Dicke state creation operator~\cite{scully2006directed}, and
\begin{subequations}
    \begin{align}
        \Psi_L^\dagger&=\sum\limits_{m=1}^{N-1} e^{i(K_\mathrm{ex}-k_0)md}L_m\sigma_m^\dagger,\\
        \Psi_R^\dagger&=\sum\limits_{m=1}^{N-1} e^{i(K_\mathrm{ex}+k_0)(N-m+1)d}R_m\sigma_{N-m+1}^\dagger    
    \end{align}
\end{subequations}
are the left- and right-edge state creation operators, respectively, with the profile factors
\begin{subequations}
    \begin{align}
        R_m &= A_\mathrm{ex}\frac{\left[z_1^\mathrm{ex}e^{i(k_0+\frac{K_\mathrm{ex}}{2})d}\right]^m}{1-z_1^\mathrm{ex}e^{i(k_0+\frac{K_\mathrm{ex}}{2})d}}+B_\mathrm{ex}\frac{\left[z_2^\mathrm{ex}e^{i(k_0+\frac{K_\mathrm{ex}}{2})d}\right]^m}{1-z_2^\mathrm{ex}e^{i(k_0+\frac{K_\mathrm{ex}}{2})d}},\\
        L_m &= A_\mathrm{ex}\frac{\left[z_1^\mathrm{ex}e^{i(k_0-\frac{K_\mathrm{ex}}{2})d}\right]^m}{1-z_1^\mathrm{ex}e^{i(k_0-\frac{K_\mathrm{ex}}{2})d}}+B_\mathrm{ex}\frac{\left[z_2^\mathrm{ex}e^{i(k_0-\frac{K_\mathrm{ex}}{2})d}\right]^m}{1-z_2^\mathrm{ex}e^{i(k_0-\frac{K_\mathrm{ex}}{2})d}}
    \end{align}
\end{subequations}
decaying exponentially with $m$. The superscript or subscript ``$\mathrm{ex}$'' denotes variables evaluated at $K_\mathrm{ex}$. Since Eq.~\eqref{eq:simplerH} has two global tails, a superposition of states $\ket*{\psi_{K_\pm}}$ with momenta $K_\pm=K_\mathrm{ex}\pm \delta K$ can be an exact eigenstate of $\mathbf{H}$ if $\mathcal{V}_{K_+}\mathcal{W}_{K_-}=\mathcal{V}_{K_-}\mathcal{W}_{K_+}$. Using the Taylor expansions $\mathbf{f}_L=f_{L1}/N+\dots$ and $\mathbf{f}_R=1+f_{R1}/N+\dots$~\cite{SM}, we find $\delta K_\zeta = (\zeta\pi/Nd)[1+(f_{R1}-f_{L1})/N]+\mathcal{O}(N^{-3})$ with $\zeta=1,2,\dots ,\zeta\ll N$. Considering the quadratic BS dispersion $\mathcal{E}_\zeta\approx\alpha_2 (\delta K_\zeta)^2$ and $\Gamma_\zeta=-2\text{Im}\mathcal{E}_\zeta$, the BS decay rate finally reads
\begin{equation}
    \Gamma_\zeta\approx\frac{\zeta^2}{N^3}|\alpha_2|\mathbf{s},
    \label{eq:BSscaling}
\end{equation}
with $\mathbf{s}=(2\pi/d)^2|\text{Im}(f_{R1}-f_{L1})|$. The corresponding BS is $\ket*{\Psi}\approx(\ket*{\psi_{K_+}}-\ket*{\psi_{K_-}})/\sqrt{2}$, which is explicitly:
\begin{equation}
    \begin{split}
        \ket*{\Psi}&=i\sqrt{\frac{2}{N}}\sum\limits_{\Delta/d=1}^{N-1}C_\Delta^\mathrm{ex}\sum\limits_{y/d=1}^{N-\Delta/d}e^{iK_\mathrm{ex}\left(y+\frac{\Delta}{2}\right)}\\
        &\quad \times \sin\left[\frac{\zeta\pi}{Nd}\left(y+\frac{\Delta}{2}\right)\right]\sigma_y^\dagger\sigma_{y+\Delta}^\dagger\ket*{G}+\mathcal{O}\left(\frac{1}{N}\right).
    \end{split}
    \label{eq:BSwf}
\end{equation}

To complete the proof, we demonstrate that the residual Hamiltonian $\Delta H=H_\mathrm{eff}-\mathbf{H}$ can be consistently treated as a perturbation~\cite{zhang2020subradiant,zhang2022free}. Specifically, we show that $\mel*{\Psi}{\Delta H}{\Psi}\sim N^{-3}$, which is an order smaller than the eigenenergy separation of $\mathbf{H}$~\cite{SM}. This validates the BS wavefunction \eqref{eq:BSwf} as the eigenstate of $H_\mathrm{eff}$.

\textit{Subradiant decay rates.\textemdash}We present numerical results of subradiant decay rates to corroborate the above theory. As shown in Fig.~\hyperref[fg:scenario]{1(c)}, we compute the decay rates of the most subradiant two-excitation eigenstates across a wide parameter range for an $N=100$ array. In the weak-chirality regime ($\eta_{LR}\lesssim1$), the most subradiant states are typically FSs, except near $d\approx \lambda_0/12$. At the spacing $d_0=\lambda_0/12$, a non-chiral array hosts a quartic BS band that renders BSs extremely subradiant states~\cite{zhang2020subradiantbound,poddubny2020quasiflat}. As chirality increases, the spacing interval over which BSs are more subradiant than FSs broadens significantly. Here, we only present results for $\eta_{LR}\geq 10^{-1.5}$. For $\eta_{LR} < 10^{-1.5}$, the BS dispersion closely approaches the continuum, leading to a strong hybridization between FSs and BSs~\cite{SM}. We thus exclude this regime to maintain an unambiguous state classification. In Fig.~\hyperref[fg:scenario]{1(d)}, we plot the infinite-array BS dispersion relations for various $\eta_{LR}$ at a fixed spacing $d=0.15\lambda_0$. We define $\mathrm{BS}_c$ and $\mathrm{BS}_e$ as the BSs located at the band-center extremum and the band-edge extremum (near the continuum), respectively. Crucially, we observe that the results in Fig.~\hyperref[fg:scenario]{1(c)} are intimately linked to the suppression of the band curvature $|\alpha_{2e}|$ for $\mathrm{BS}_e$.

To further elucidate the mechanism underlying this BS subradiance enhancement, we explicitly show the impact of chirality on the decay rates in Fig.~\hyperref[fg:BSeAlpha2]{2(a)}. Notably, as $\eta_{LR}$ is reduced, the decay rate of $\mathrm{BS}_e$ (with $\zeta=1$ by default) first decreases and then increases. As demonstrated in Fig.~\hyperref[fg:BSeAlpha2]{2(b)}, this non-monotonic behavior stems directly from the variation in $|\alpha_{2e}|$. Physically, the introduction of chiral interactions breaks the parity symmetry of the two-excitation dispersion. This leads to a reorganization of the continuum, where the maximum of the lower branch shifts rightward and upward, while the lower boundary of the upper branch descends, effectively narrowing the bandgap that hosts the BS dispersion. The extremum associated with $\mathrm{BS}_e$ shifts correspondingly, and its curvature $|\alpha_{2e}|$ can be suppressed for $\eta_{LR}\lesssim 1$ (see Fig.~\hyperref[fg:scenario]{1(d)} and Fig.~S5 in Ref.~\cite{SM}). As implied by the numerical results, the dependence of the $\mathrm{BS}_e$ decay rate on $\eta_{LR}$ is similar to that of $|\alpha_{2e}|$. To further show that the subradiant decay rate is dominated by $|\alpha_2|$, according to Eq.~(\ref{eq:BSscaling}), we extract the value of $\mathbf{s}_e$ using $\mathbf{s}_e=\Gamma N^3/|\alpha_{2e}|$, which exhibits a much gentler variation with $\eta_{LR}$ compared to $|\alpha_{2e}|$. Conversely, the decay rates of both $\mathrm{BS}_c$ and the FS display an overall increase, with $\mathrm{BS}_c$ exhibiting oscillatory behavior during its rise. Consequently, under sufficiently strong chirality, both become less subradiant than $\mathrm{BS}_e$.

\begin{figure}[htbp]
    \includegraphics[scale=0.54]{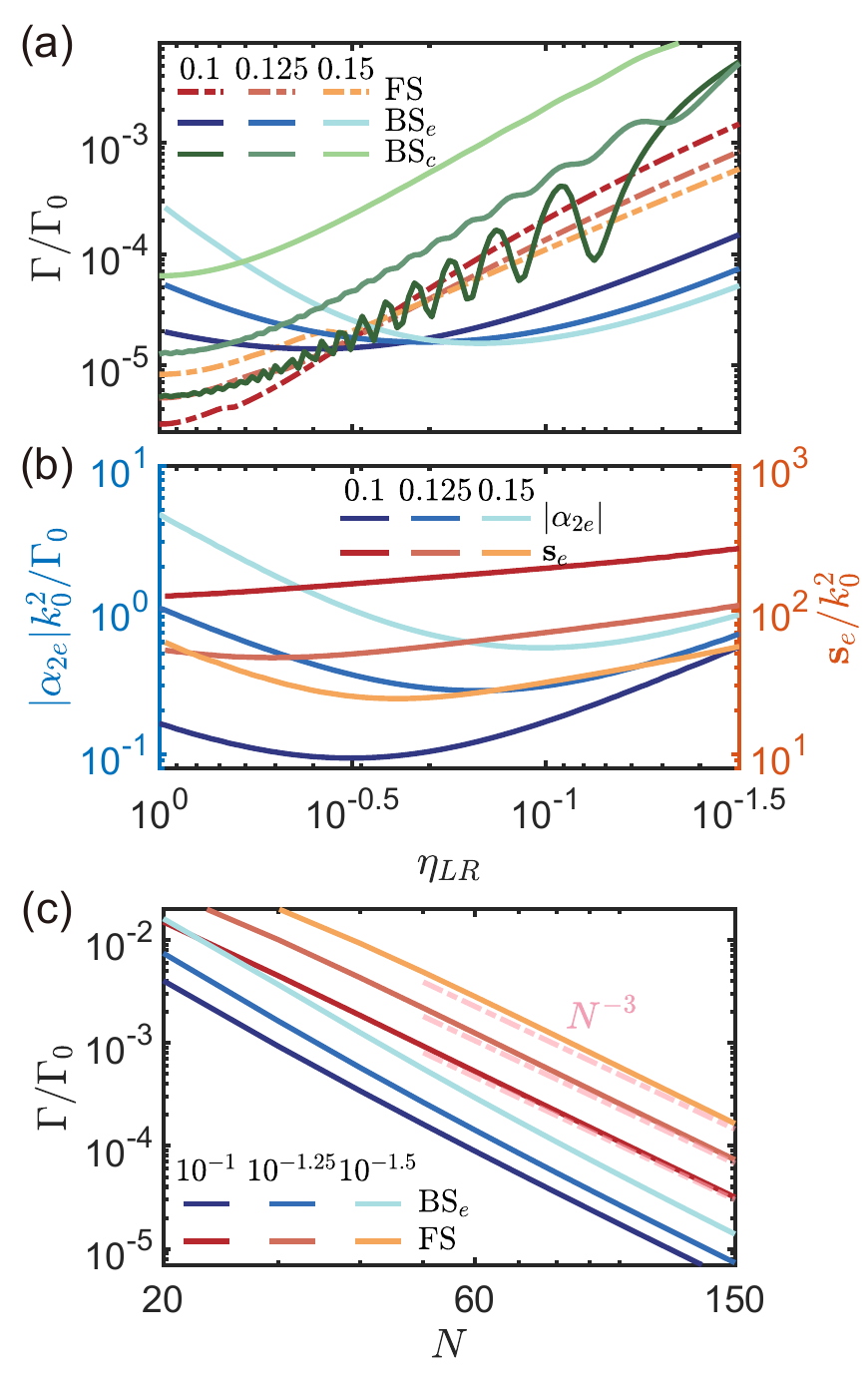}
    \centering
    \caption{(a) Decay rates of $\mathrm{FS}$ (red), $\mathrm{BS}_e$ (blue), and $\mathrm{BS}_c$ (green) as a function of $\eta_{LR}$ with $N=100$, where the values of $d/\lambda_0$ are specified in the legend.
    (b) Dependence of $|\alpha_{2e}|$ (blue) and $\mathbf{s}_e$ (red) on $\eta_{LR}$, with $d/\lambda_0$ indicated by the legend.
    The extraction of $\mathbf{s}_e$ utilizes the $\mathrm{BS}_e$ results from (a) for $N=100$.
    (c) Decay rate scaling of $\mathrm{BS}_e$ (blue) and $\mathrm{FS}$ (red) with respect to the atom number $N$ for three different $\eta_{LR}$ values at a fixed spacing $d=0.15\lambda_0$. The dash-dotted pink lines denote the analytical scaling of the FS decay rates.} 
    \label{fg:BSeAlpha2} 
\end{figure}

Finally, the numerical results in Fig.~\hyperref[fg:BSeAlpha2]{2(c)} (evaluated at $d=0.15\lambda_0$) confirm that the BS decay rates follow the $\Gamma\sim N^{-3}$ scaling for sufficiently large $N$, which coincides with Eq.~\eqref{eq:BSscaling}. For comparison, we also show the decay rate scaling of FSs, along with the analytical expression $\Gamma_{\bm{\xi}}=\sum_{s=1}^{n_e} \gamma_{\xi_s}$ with~\cite{SM}
\begin{equation}
    \begin{split}
        \gamma_\xi&\approx\frac{\pi^2\xi^2}{8N^3}\frac{\sin k_0d}{\sin\frac{(k_0+k_\mathrm{ex})d}{2}\sin\frac{(k_0-k_\mathrm{ex})d}{2}}\\&\quad\times\left[\Gamma_L\frac{\cot\frac{(k_0+k_\mathrm{ex})d}{2}}{\sin^2\frac{(k_0+k_\mathrm{ex})d}{2}}+\Gamma_R\frac{\cot\frac{(k_0-k_\mathrm{ex})d}{2}}{\sin^2\frac{(k_0-k_\mathrm{ex})d}{2}}\right].
    \end{split}
    \label{eq:singleScaling}
\end{equation}
Here, $k_\text{ex}=k_0-(2/d)\cot^{-1}[(\cos k_0d\pm\sqrt{\eta_{LR}})/\sin k_0d]$ is the extremum point of the single-excitation band. The most subradiant two-excitation FS has excitation number $n_e=2$ and fermionic indices $\bm{\xi}=(1,2)$. Furthermore, we demonstrate the $\zeta^2$-dependence of the BS decay rates in the Supplemental Material~\cite{SM}. 

We anticipate that the modulation of subradiance by chirality is universal. To this end, we introduce a simpler toy model in the Supplemental Material~\cite{SM}, demonstrating that the reduction of $|\alpha_2|$ due to chirality is intimately tied to the suppression of subradiant decay rates.

\begin{figure}[htbp]
    \includegraphics[scale=0.51]{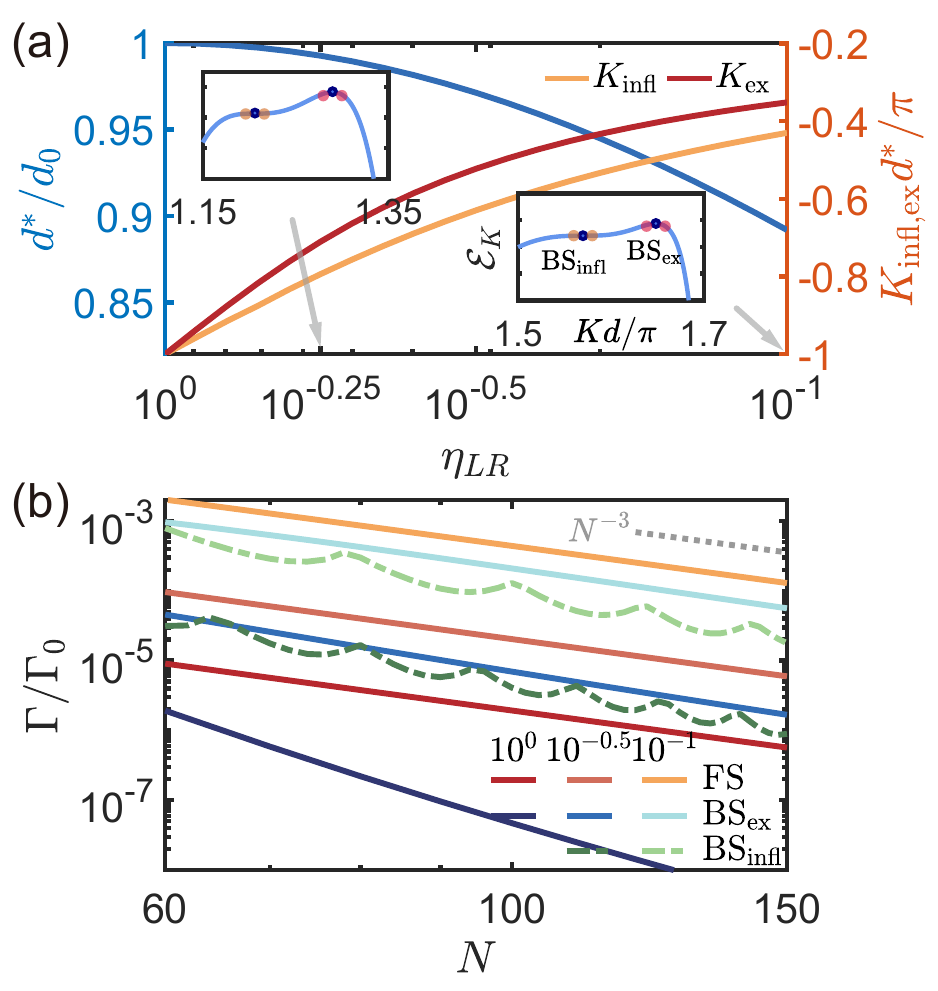}
    \centering
    \caption{(a) Blue solid curve is the critical spacing where the cubic band emerges for a given chirality. The inflection point and extremum point are represented by the yellow and red curves (shifted into the first Brillouin zone), respectively. Insets show the BS dispersion relations for $\eta_{LR}=10^{-0.25}$ and $\eta_{LR}=10^{-1}$, where the circles mark $\mathrm{BS}_\mathrm{infl}$ (yellow) and $\mathrm{BS}_\mathrm{ex}$ (red) for finite arrays with $N=100$.
    (b) Decay rate scaling of $\mathrm{FS}$, $\mathrm{BS}_\mathrm{ex}$, and $\mathrm{BS}_\mathrm{infl}$ for $\eta_{LR}=1$, $10^{-0.5}$ and $10^{-1}$ (indicated by the legend), evaluated at the critical spacing $d=d^*$ for each chirality. The dotted line provides a reference for the $N^{-3}$ dependence.} 
    \label{fg:BSinfl} 
\end{figure}

\textit{Oscillating decay rate.\textemdash}In addition, we observe a subradiant dip in Fig.~\hyperref[fg:scenario]{1(c)}. The dip starts from the point $d=d_0$ for $\eta_{LR}=1$, indicating the presence of an extraordinary BS dispersion. To characterize this quantitatively, we extract the corresponding BS band in Fig.~\hyperref[fg:BSinfl]{3(a)}. For $\eta_{LR}<1$, a cubic dispersion $\mathcal{E}_K\approx \alpha_3(K-K_\mathrm{infl})^3$ emerges near the inflection point $K_\mathrm{infl}$ when the lattice reaches a critical spacing $d=d^*$. This gives rise to a novel class of BSs, denoted as $\mathrm{BS}_\mathrm{infl}$, located in the vicinity of $K_\mathrm{infl}$ for finite arrays. Interestingly, as revealed in Fig.~\hyperref[fg:BSinfl]{3(b)}, $\mathrm{BS}_\mathrm{infl}$ exhibits an oscillating decay rate as a function of $N$, contrasting with the $N^{-3}$ scaling observed for BSs near quadratic extrema (denoted here as $\mathrm{BS}_\mathrm{ex}$). We attribute this oscillation to the critical nature of the cubic dispersion. When $d>d^*$, the BS band has three distinct extrema, each mediating an $N^{-3}$ scaling. However, as $d$ decreases past $d^*$, the two left extrema merge, resulting in a locally linear dispersion near the inflection point. Such a linear dispersion suggests an $N^{-1}$ scaling, considering the argument $\zeta\sim \mathcal{O}(N)$~\cite{asenjo2017exponential,zhang2020subradiant}. Consequently, the critical spacing $d=d^*$ hosts a hybridized scaling that is manifested as an oscillation with $N$. Similar oscillatory behaviors induced by specific lattice geometry have been noted in Refs.~\cite{zhang2020subradiantbound,poddubny2020quasiflat,zhang2020subradiant}.

\begin{figure}[htbp]
    \includegraphics[scale=0.59]{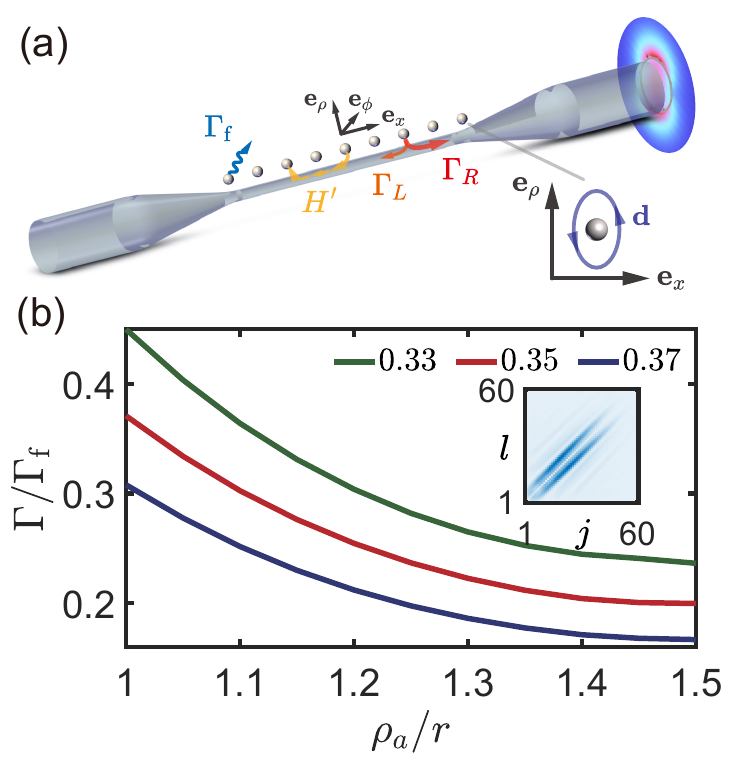}
    \centering
    \caption{(a) Schematic of an atom array coupled to a cylindrical optical nanofiber.
    The coupling between the elliptical atomic dipoles and the spin-momentum locked guided mode leads to chiral emission. (b) Decay rates of the chiral BSs for an $N=60$ array, with $d/\lambda_\mathrm{f}=0.33$ (green), $0.35$ (red), and $0.37$ (blue).
    Inset: probability distribution of a subradiant chiral BS, illustrated for parameters $d=0.35\lambda_\mathrm{f}$, $\rho_a=r$, and $N=60$.} 
    \label{fg:BSONF} 
\end{figure}

\textit{Nanofiber interface.\textemdash}To demonstrate the physical realization of these subradiant chiral BSs on realistic experimental platforms, we investigate an $N$-atom array coupled to a single-mode optical nanofiber. As illustrated in Fig.~\hyperref[fg:BSONF]{4(a)}, the total atom-atom interaction is governed by the Hamiltonian~\cite{le2017nanofiber}
\begin{equation}
    H=H_\mathrm{eff}+H'.\label{eq:ONF}
\end{equation}
Here, $H_\mathrm{eff}$ represents the chiral interaction mediated by the guided $\text{HE}_{11}$ mode, retaining the form of Eq.~\eqref{eq:Heff}, while $H'$ accounts for the interaction via non-guided radiation modes. The chirality is directly related to the atomic dipole moment $\mathbf{d}$, which asymmetrically couples to the left- and right-propagating guided modes when the atoms are not linearly polarized. 

In our numerical calculations, we consider a silica nanofiber in vacuum with a core refractive index of $n=1.45$ and a radius of $r=180\,\text{nm}$. The emitters are modeled on the $\text{D}_2$ transition of $^{87}\text{Rb}$, with a resonant frequency $\omega_\mathrm{f}=2\pi\times384.23\,\text{THz}$ and a natural linewidth $\Gamma_\mathrm{f}=2\pi\times6.07\,\text{MHz}$~\cite{Steck2003}. This configuration yields a guided wavevector $k_0=1.06\times 2\pi/\lambda_\mathrm{f}$, where $\lambda_\mathrm{f}=2\pi c/\omega_\mathrm{f}$ is the free-space wavelength. By diagonalizing $H$ in the two-excitation manifold, we identify the subradiant BSs, as depicted in the inset of Fig.~\hyperref[fg:BSONF]{4(b)}. For this demonstration, we consider a transition dipole moment $\mathbf{d}\propto (3/2)\mathbf{e}_\rho+i\mathbf{e}_x$, which can be created, e.g., by dressing the excited state with linearly polarized light. This choice of polarization helps to accentuate the characteristics of the BSs. We show in the Supplemental Material~\cite{SM} that the left-concentrated BS spatial profile, reminiscent of the non-Hermitian skin effect in Ref.~\cite{shi2025chiral}, indeed, arises from the chiral interaction. The decay rates plotted against the atomic radial coordinate $\rho_a$ in Fig.~\hyperref[fg:BSONF]{4(b)} further confirm the subradiant property of these states. Notably, the BSs at the ONF interface exhibit features of hybridization with scattering resonances, which likely arise from the non-guided interaction channel $H'$ in Eq.~\eqref{eq:ONF}.

\textit{Conclusions.\textemdash} We have derived analytically that the decay rate of subradiant BSs in chiral waveguide QED scales as $\Gamma\sim|\alpha_2|/N^3$. This formulation clarifies that chirality-induced suppression of the band curvature, $|\alpha_2|$, is the driving mechanism that allows BSs to become the most subradiant two-excitation eigenstates across a broad range of lattice spacings. We also identified a novel class of BSs located near the inflection point of a cubic dispersion band with vanishing $\alpha_2$, which are characterized by a size-dependent oscillating decay rate. We have also confirmed the existence of the subradiant chiral BSs in realistic optical nanofiber interfaces, despite the strong non-guided dipole interactions therein.

It is known that the subradiance of 1D atom arrays can be engineered via polaritonic dispersion~\cite{zhang2020subradiant,poddubny2020quasiflat,zhang2022free,tevcer2026flat}. In this work, we show that chirality provides a powerful and highly tunable degree of freedom to engineer these band structures in the multi-excitation regime. Since chiral light-matter interactions exist ubiquitously in nanophotonic systems~\cite{lodahl2017chiral,suarez2025chiral}, we anticipate that the subradiance mechanism revealed in this work can be generalized far beyond standard chiral waveguide QED setups. Exploring the impact of chirality on
many-body physics in higher-dimensional atom arrays~\cite{marques2021bound,tevcer2024strongly,calajo2025many,tevcer2026flat} or arrays coupled to structured photonic reservoirs~\cite{bello2019unconventional,perczel2020topological,vega2023topological,tian2024power} may be a promising future avenue. 

Experimentally, the excitation of chiral BSs in nanofiber setups could be realized via solitonic pulse schemes~\cite{calajo2022emergence}. To this end, developing error-resilient multiphoton control techniques to mitigate experimental imperfections, such as atomic thermal motion~\cite{rusconi2021exploiting,castells2025cavity} or quantum vibration~\cite{olmos2025hybrid,eltohfa2025effects}, would be crucial for practical implementations. Moreover, exploiting this paradigm on chiral superconducting circuit platforms~\cite{zhang2021charge,wang2022chiral} represents a promising frontier. Further applications, including robust photon-pair storage and directional transport, also warrant future investigation.

\vspace{10pt}
\textit{Acknowledgments.\textemdash}We thank Zijian Du for providing computational resources. K. T. acknowledges Professor Yu-Xiang Zhang and Professor Darrick Chang for their theoretical guidance. This project is supported by National Key Research Program of China under Grant No. 2022YFA1404204, from Natural Science Foundation of Shanghai
under Grant No. 23dz2260100, and from the Shanghai Science and Technology Innovation Action Plan under the Grant No. 24LZ1400300.

\nocite{*}
\bibliography{main}

@Article{shah2024quantum,
  author    = {Shah, Freya and Patti, Taylor L and Rubies-Bigorda, Oriol and Yelin, Susanne F},
  title     = {Quantum computing with subwavelength atomic arrays},
  doi       = {10.1103/physreva.109.012613},
  number    = {1},
  pages     = {012613},
  volume    = {109},
  journal   = {Phys. Rev. A},
  publisher = {APS},
  year      = {2024},
}

@Article{gonzalez2024light,
  author    = {Gonz{\'a}lez-Tudela, Alejandro and Reiserer, Andreas and Garc{\'\i}a-Ripoll, Juan Jos{\'e} and Garc{\'\i}a-Vidal, Francisco J},
  title     = {Light--matter interactions in quantum nanophotonic devices},
  doi       = {10.1038/s42254-023-00681-1},
  number    = {3},
  pages     = {166--179},
  volume    = {6},
  journal   = {Nat. Rev. Phys.},
  publisher = {Nature Publishing Group UK London},
  year      = {2024},
}

@Article{sheremet2023waveguide,
  author    = {Sheremet, Alexandra S and Petrov, Mihail I and Iorsh, Ivan V and Poshakinskiy, Alexander V and Poddubny, Alexander N},
  title     = {Waveguide quantum electrodynamics: Collective radiance and photon-photon correlations},
  doi       = {https://doi.org/10.1103/revmodphys.95.015002},
  number    = {1},
  pages     = {015002},
  volume    = {95},
  journal   = {Rev. Mod. Phys.},
  publisher = {APS},
  year      = {2023},
}

@Article{dhordjevic2021entanglement,
  author    = {{\DH}or{\dj}evi{\'c}, Tamara and Samutpraphoot, Polnop and Ocola, Paloma L and Bernien, Hannes and Grinkemeyer, Brandon and Dimitrova, Ivana and Vuleti{\'c}, Vladan and Lukin, Mikhail D},
  title     = {Entanglement transport and a nanophotonic interface for atoms in optical tweezers},
  doi       = {10.1126/science.abi9917},
  number    = {6562},
  pages     = {1511--1514},
  volume    = {373},
  journal   = {Science},
  publisher = {American Association for the Advancement of Science},
  year      = {2021},
}

@Article{zhang2022free,
  author    = {Zhang, Yu-Xiang and M{\o}lmer, Klaus},
  title     = {Free-fermion multiply excited eigenstates and their experimental signatures in 1d arrays of two-level atoms},
  doi       = {https://doi.org/10.1103/PhysRevLett.128.093602},
  number    = {9},
  pages     = {093602},
  volume    = {128},
  journal   = {Phys. Rev. Lett.},
  publisher = {APS},
  year      = {2022},
}

@Article{lodahl2017chiral,
  author    = {Lodahl, Peter and Mahmoodian, Sahand and Stobbe, S{\o}ren and Rauschenbeutel, Arno and Schneeweiss, Philipp and Volz, J{\"u}rgen and Pichler, Hannes and Zoller, Peter},
  title     = {Chiral quantum optics},
  doi       = {10.1038/nature21037},
  number    = {7638},
  pages     = {473--480},
  volume    = {541},
  journal   = {Nature},
  publisher = {Nature Publishing Group UK London},
  year      = {2017},
}

@Article{mahmoodian2020dynamics,
  author    = {Mahmoodian, Sahand and Calaj{\'o}, Giuseppe and Chang, Darrick E and Hammerer, Klemens and S{\o}rensen, Anders S},
  title     = {Dynamics of many-body photon bound states in chiral waveguide QED},
  doi       = {https://doi.org/10.1103/PhysRevX.10.031011},
  number    = {3},
  pages     = {031011},
  volume    = {10},
  journal   = {Phys. Rev. X},
  publisher = {APS},
  year      = {2020},
}

@Article{bakkensen2021photonic,
  author  = {Bakkensen, Bastian and Zhang, Yu-Xiang and Bjerlin, Johannes and S{\o}rensen, Anders S{\o}ndberg},
  title   = {Photonic bound states and scattering resonances in waveguide QED},
  url     = {https://doi.org/10.48550/arXiv.2110.06093},
  journal = {arXiv:2110.06093},
  year    = {2021},
}

@Article{calajo2025many,
  author    = {Calaj{\'o}, Giuseppe and Te{\v{c}}er, Matija and Montangero, Simone and Silvi, Pietro and Di Liberto, Marco},
  title     = {Many-body quantum dimerization in two-dimensional atomic arrays},
  doi       = {10.1103/1r6c-6ftv},
  number    = {1},
  pages     = {013722},
  volume    = {112},
  journal   = {Phys. Rev. A},
  publisher = {APS},
  year      = {2025},
}

@Article{perczel2020topological,
  author    = {Perczel, Janos and Borregaard, Johannes and Chang, Darrick E and Yelin, Susanne F and Lukin, Mikhail D},
  title     = {Topological quantum optics using atomlike emitter arrays coupled to photonic crystals},
  doi       = {https://doi.org/10.1103/PhysRevLett.124.083603},
  number    = {8},
  pages     = {083603},
  volume    = {124},
  journal   = {Phys. Rev. Lett.},
  publisher = {APS},
  year      = {2020},
}

@Article{asenjo2017exponential,
  author    = {Asenjo-Garcia, Ana and Moreno-Cardoner, M and Albrecht, Andreas and Kimble, HJ and Chang, Darrick E},
  title     = {Exponential improvement in photon storage fidelities using subradiance and “selective radiance” in atomic arrays},
  doi       = {10.1103/physrevx.7.031024},
  number    = {3},
  pages     = {031024},
  volume    = {7},
  journal   = {Phys. Rev. X},
  publisher = {APS},
  year      = {2017},
}

@Article{tevcer2026flat,
  author    = {Te{\v{c}}er, Matija and Calaj{\'o}, Giuseppe and Di Liberto, Marco},
  title     = {Flat-band-mediated photon-photon interactions in two-dimensional waveguide QED networks},
  doi       = {10.1103/dt14-h2c2},
  number    = {1},
  pages     = {013701},
  volume    = {113},
  journal   = {Phys. Rev. A},
  publisher = {APS},
  year      = {2026},
}

@Article{le2017nanofiber,
  author    = {Le Kien, Fam and Rauschenbeutel, Arno},
  title     = {Nanofiber-mediated chiral radiative coupling between two atoms},
  doi       = {10.1103/physreva.95.023838},
  number    = {2},
  pages     = {023838},
  volume    = {95},
  journal   = {Phys. Rev. A},
  publisher = {APS},
  year      = {2017},
}

@Article{liedl2024observation,
  author    = {Liedl, Christian and Tebbenjohanns, Felix and Bach, Constanze and Pucher, Sebastian and Rauschenbeutel, Arno and Schneeweiss, Philipp},
  title     = {Observation of superradiant bursts in a cascaded quantum system},
  doi       = {10.1103/physrevx.14.011020},
  number    = {1},
  pages     = {011020},
  volume    = {14},
  journal   = {Phys. Rev. X},
  publisher = {APS},
  year      = {2024},
}

@Article{schrinski2022polariton,
  author    = {Schrinski, Bj{\"o}rn and S{\o}rensen, Anders S},
  title     = {Polariton dynamics in one-dimensional arrays of atoms coupled to waveguides},
  doi       = {10.1088/1367-2630/acaa4f},
  number    = {12},
  pages     = {123023},
  volume    = {24},
  journal   = {New J. Phys.},
  publisher = {IOP Publishing},
  year      = {2022},
}

@Article{chang2012cavity,
  author    = {Chang, Darrick E and Jiang, L and Gorshkov, AV and Kimble, HJ},
  title     = {Cavity QED with atomic mirrors},
  doi       = {10.1088/1367-2630/14/6/063003},
  number    = {6},
  pages     = {063003},
  volume    = {14},
  journal   = {New J. Phys.},
  publisher = {IOP Publishing},
  year      = {2012},
}

@Article{scully2006directed,
  author    = {Scully, Marlan O and Fry, Edward S and Ooi, CH Raymond and W{\'o}dkiewicz, Krzysztof},
  title     = {Directed spontaneous emission from an extended ensemble of N atoms: timing is everything},
  doi       = {https://doi.org/10.1103/PhysRevLett.96.010501},
  number    = {1},
  pages     = {010501},
  volume    = {96},
  journal   = {Phys. Rev. Lett.},
  publisher = {APS},
  year      = {2006},
}

@Article{lu2025chiral,
  author    = {Lu, Zhi-Guang and Wu, Ying and L{\"u}, Xin-You},
  title     = {Chiral interaction induced near-perfect photon blockade},
  doi       = {https://doi.org/10.1103/PhysRevLett.134.013602},
  number    = {1},
  pages     = {013602},
  volume    = {134},
  journal   = {Phys. Rev. Lett.},
  publisher = {APS},
  year      = {2025},
}

@Article{dicke1954coherence,
  author    = {Dicke, Robert H},
  title     = {Coherence in spontaneous radiation processes},
  doi       = {10.1103/physrev.93.99},
  number    = {1},
  pages     = {99},
  volume    = {93},
  journal   = {Phys. Rev.},
  publisher = {APS},
  year      = {1954},
}

@Article{moore2025van,
  author    = {Moore, Samuel L and Lee, Hae Yeon and Rivera, Nicholas and Karube, Yuzuka and Ziffer, Mark and Yanev, Emanuil S and Darlington, Thomas P and Sternbach, Aaron J and Holbrook, Madisen A and Pack, Jordan and others},
  title     = {Van der Waals waveguide quantum electrodynamics probed by infrared nano-photoluminescence},
  doi       = {10.1038/s41566-025-01694-1},
  number    = {8},
  pages     = {833--839},
  volume    = {19},
  journal   = {Nat. Photonics},
  publisher = {Nature Publishing Group UK London},
  year      = {2025},
}

@Article{corzo2016large,
  author    = {Corzo, Neil V and Gouraud, Baptiste and Chandra, Aveek and Goban, Akihisa and Sheremet, Alexandra S and Kupriyanov, Dmitriy V and Laurat, Julien},
  title     = {Large Bragg reflection from one-dimensional chains of trapped atoms near a nanoscale waveguide},
  doi       = {10.1103/physrevlett.117.133603},
  number    = {13},
  pages     = {133603},
  volume    = {117},
  journal   = {Phys. Rev. Lett.},
  publisher = {APS},
  year      = {2016},
}

@Article{bach2026emergence,
  author    = {Bach, Constanze and Tebbenjohanns, Felix and Liedl, Christian and Schneeweiss, Philipp and Rauschenbeutel, Arno},
  title     = {Emergence of second-order coherence in superfluorescence},
  doi       = {10.1103/r3kg-c4x9},
  number    = {6},
  pages     = {063402},
  volume    = {136},
  journal   = {Phys. Rev. Lett.},
  publisher = {APS},
  year      = {2026},
}

@Article{suarez2025chiral,
  author    = {Su{\'a}rez-Forero, DG and Jalali Mehrabad, M and Vega, C and Gonz{\'a}lez-Tudela, A and Hafezi, M},
  title     = {Chiral quantum optics: recent developments and future directions},
  doi       = {10.1103/prxquantum.6.020101},
  number    = {2},
  pages     = {020101},
  volume    = {6},
  journal   = {PRX Quantum},
  publisher = {APS},
  year      = {2025},
}

@Article{olmos2025hybrid,
  author    = {Olmos, Beatriz and Lesanovsky, Igor},
  title     = {Hybrid sub-and superradiant states in emitter arrays with quantized motion},
  doi       = {https://doi.org/10.1103/q2kj-w3lf},
  number    = {24},
  pages     = {243602},
  volume    = {134},
  journal   = {Phys. Rev. Lett.},
  publisher = {APS},
  year      = {2025},
}

@Article{zhang2020subradiantbound,
  author    = {Zhang, Yu-Xiang and Yu, Chuan and M{\o}lmer, Klaus},
  title     = {Subradiant bound dimer excited states of emitter chains coupled to a one dimensional waveguide},
  doi       = {https://doi.org/10.1103/PhysRevResearch.2.013173},
  number    = {1},
  pages     = {013173},
  volume    = {2},
  journal   = {Phys. Rev. Res.},
  publisher = {APS},
  year      = {2020},
}

@Article{zhang2019theory,
  author    = {Zhang, Yu-Xiang and M{\o}lmer, Klaus},
  title     = {Theory of subradiant states of a one-dimensional two-level atom chain},
  doi       = {https://doi.org/10.1103/PhysRevLett.122.203605},
  number    = {20},
  pages     = {203605},
  volume    = {122},
  journal   = {Phys. Rev. Lett.},
  publisher = {APS},
  year      = {2019},
}

@Article{pichler2015quantum,
  author    = {Pichler, Hannes and Ramos, Tom{\'a}s and Daley, Andrew J and Zoller, Peter},
  title     = {Quantum optics of chiral spin networks},
  doi       = {10.1103/physreva.91.042116},
  number    = {4},
  pages     = {042116},
  volume    = {91},
  journal   = {Phys. Rev. A},
  publisher = {APS},
  year      = {2015},
}

@Article{moreno2021quantum,
  author    = {Moreno-Cardoner, Mariona and Goncalves, Daniel and Chang, Darrick E},
  title     = {Quantum nonlinear optics based on two-dimensional Rydberg atom arrays},
  doi       = {https://doi.org/10.1103/PhysRevLett.127.263602},
  number    = {26},
  pages     = {263602},
  volume    = {127},
  journal   = {Phys. Rev. Lett.},
  publisher = {APS},
  year      = {2021},
}

@Article{chang2018colloquium,
  author    = {Chang, DE and Douglas, JS and Gonz{\'a}lez-Tudela, Alejandro and Hung, C-L and Kimble, HJ},
  title     = {Colloquium: Quantum matter built from nanoscopic lattices of atoms and photons},
  doi       = {https://doi.org/10.1103/RevModPhys.90.031002},
  number    = {3},
  pages     = {031002},
  volume    = {90},
  journal   = {Rev. Mod. Phys.},
  publisher = {APS},
  year      = {2018},
}

@Article{poddubny2020quasiflat,
  author    = {Poddubny, Alexander N},
  title     = {Quasiflat band enabling subradiant two-photon bound states},
  doi       = {10.1103/physreva.101.043845},
  number    = {4},
  pages     = {043845},
  volume    = {101},
  journal   = {Phys. Rev. A},
  publisher = {APS},
  year      = {2020},
}

@article{jones2020collectively,
  title={Collectively enhanced chiral photon emission from an atomic array near a nanofiber},
  author={Jones, Ryan and Buonaiuto, Giuseppe and Lang, Ben and Lesanovsky, Igor and Olmos, Beatriz},
  journal={Phys. Rev. Lett.},
  volume={124},
  number={9},
  pages={093601},
  year={2020},
  publisher={APS},
  doi={https://doi.org/10.1103/PhysRevLett.124.093601}
}

@Article{albrecht2019subradiant,
  author    = {Albrecht, Andreas and Henriet, Lo{\"\i}c and Asenjo-Garcia, Ana and Dieterle, Paul B and Painter, Oskar and Chang, Darrick E},
  title     = {Subradiant states of quantum bits coupled to a one-dimensional waveguide},
  doi       = {https://doi.org/10.1088/1367-2630/ab0134},
  number    = {2},
  pages     = {025003},
  volume    = {21},
  journal   = {New J. Phys.},
  publisher = {IOP Publishing},
  year      = {2019},
}

@Article{rui2020subradiant,
  author    = {Rui, Jun and Wei, David and Rubio-Abadal, Antonio and Hollerith, Simon and Zeiher, Johannes and Stamper-Kurn, Dan M and Gross, Christian and Bloch, Immanuel},
  title     = {A subradiant optical mirror formed by a single structured atomic layer},
  doi       = {10.1038/s41586-020-2463-x},
  number    = {7816},
  pages     = {369--374},
  volume    = {583},
  journal   = {Nature},
  publisher = {Nature Publishing Group UK London},
  year      = {2020},
}

@Article{prasad2020correlating,
  author    = {Prasad, Adarsh S and Hinney, Jakob and Mahmoodian, Sahand and Hammerer, Klemens and Rind, Samuel and Schneeweiss, Philipp and S{\o}rensen, Anders S and Volz, J{\"u}rgen and Rauschenbeutel, Arno},
  title     = {Correlating photons using the collective nonlinear response of atoms weakly coupled to an optical mode},
  doi       = {https://doi.org/10.1038/s41566-020-0692-z},
  number    = {12},
  pages     = {719--722},
  volume    = {14},
  journal   = {Nat. Photonics},
  publisher = {Nature Publishing Group UK London},
  year      = {2020},
}

@Article{tabares2023variational,
  author    = {Tabares, Cristian and Mu{\~n}oz de Las Heras, Alberto and Tagliacozzo, Luca and Porras, Diego and Gonz{\'a}lez-Tudela, Alejandro},
  title     = {Variational quantum simulators based on waveguide QED},
  doi       = {10.1103/physrevlett.131.073602},
  number    = {7},
  pages     = {073602},
  volume    = {131},
  journal   = {Phys. Rev. Lett.},
  publisher = {APS},
  year      = {2023},
}

@Article{chang2014quantum,
  author    = {Chang, Darrick E and Vuleti{\'c}, Vladan and Lukin, Mikhail D},
  title     = {Quantum nonlinear optics—photon by photon},
  doi       = {10.1038/nphoton.2014.192},
  number    = {9},
  pages     = {685--694},
  volume    = {8},
  journal   = {Nat. Photonics},
  publisher = {Nature Publishing Group UK London},
  year      = {2014},
}

@Article{tomas2025passive,
  author    = {Levy-Yeyati, Tom\'as and Vega, Carlos and Ramos, Tom\'as and Gonz\'alez-Tudela, Alejandro},
  title     = {Passive Photonic CZ Gate with Two-Level Emitters in Chiral Multimode Waveguide QED},
  doi       = {10.1103/PRXQuantum.6.010342},
  issue     = {1},
  pages     = {010342},
  url       = {https://link.aps.org/doi/10.1103/PRXQuantum.6.010342},
  volume    = {6},
  journal   = {PRX Quantum},
  month     = {Mar},
  numpages  = {18},
  publisher = {American Physical Society},
  year      = {2025},
}

@Article{sunami2025scalable,
  author    = {Sunami, Shinichi and Tamiya, Shiro and Inoue, Ryotaro and Yamasaki, Hayata and Goban, Akihisa},
  title     = {Scalable networking of neutral-atom qubits: Nanofiber-based approach for multiprocessor fault-tolerant quantum computers},
  doi       = {10.1103/prxquantum.6.010101},
  number    = {1},
  pages     = {010101},
  volume    = {6},
  journal   = {PRX Quantum},
  publisher = {APS},
  year      = {2025},
}

@Article{vega2023topological,
  author    = {Vega, Carlos and Porras, Diego and Gonz{\'a}lez-Tudela, Alejandro},
  title     = {Topological multimode waveguide QED},
  doi       = {10.1103/physrevresearch.5.023031},
  number    = {2},
  pages     = {023031},
  volume    = {5},
  journal   = {Phys. Rev. Res.},
  publisher = {APS},
  year      = {2023},
}

@Article{Steck2003,
  author  = {Steck, Daniel A},
  title   = {{Rubidium 87 D line data (Version 2.3.4, last revised 8 August 2025)}},
  url     = {http://steck.us/alkalidata},
  journal = {http://steck.us/alkalidata},
}

@Article{perczel2017topological,
  author    = {Perczel, Janos and Borregaard, Johannes and Chang, Darrick E and Pichler, Hannes and Yelin, Susanne F and Zoller, Peter and Lukin, Mikhail D},
  title     = {Topological quantum optics in two-dimensional atomic arrays},
  doi       = {https://doi.org/10.1103/PhysRevLett.119.023603},
  number    = {2},
  pages     = {023603},
  volume    = {119},
  journal   = {Phys. Rev. Lett.},
  publisher = {APS},
  year      = {2017},
}

@Article{asenjo2017atom,
  author    = {Asenjo-Garcia, Ana and Hood, JD and Chang, DE and Kimble, HJ},
  title     = {Atom-light interactions in quasi-one-dimensional nanostructures: A Green's-function perspective},
  doi       = {10.1103/physreva.95.033818},
  number    = {3},
  pages     = {033818},
  volume    = {95},
  journal   = {Phys. Rev. A},
  publisher = {APS},
  year      = {2017},
}

@Article{SM,
  title   = {{See Supplemental Material for detailed derivations, additional plots and extended discussions, which includes Refs.~\cite{fedorovich2022chirality,zhong2020photon,le2005nanofiber,tong2004single,solano2017optical,svendsen2023modified}}},
  journal = {..},
}

@Article{tevcer2024strongly,
  author    = {Te{\v{c}}er, Matija and Di Liberto, Marco and Silvi, Pietro and Montangero, Simone and Romanato, Filippo and Calaj{\'o}, Giuseppe},
  title     = {Strongly interacting photons in 2D waveguide QED},
  doi       = {10.1103/physrevlett.132.163602},
  number    = {16},
  pages     = {163602},
  volume    = {132},
  journal   = {Phys. Rev. Lett.},
  publisher = {APS},
  year      = {2024},
}

@Article{paulisch2016universal,
  author    = {Paulisch, Vanessa and Kimble, HJ and Gonz{\'a}lez-Tudela, Alejandro},
  title     = {Universal quantum computation in waveguide QED using decoherence free subspaces},
  doi       = {10.1088/1367-2630/18/4/043041},
  number    = {4},
  pages     = {043041},
  volume    = {18},
  journal   = {New J. Phys.},
  publisher = {IOP Publishing},
  year      = {2016},
}

@Article{corzo2019waveguide,
  author    = {Corzo, Neil V and Raskop, J{\'e}r{\'e}my and Chandra, Aveek and Sheremet, Alexandra S and Gouraud, Baptiste and Laurat, Julien},
  title     = {Waveguide-coupled single collective excitation of atomic arrays},
  doi       = {10.1038/s41586-019-0902-3},
  number    = {7744},
  pages     = {359--362},
  volume    = {566},
  journal   = {Nature},
  publisher = {Nature Publishing Group UK London},
  year      = {2019},
}

@Article{henriet2019critical,
  author    = {Henriet, Lo{\"\i}c and Douglas, James S and Chang, Darrick E and Albrecht, Andreas},
  title     = {Critical open-system dynamics in a one-dimensional optical-lattice clock},
  doi       = {10.1103/physreva.99.023802},
  number    = {2},
  pages     = {023802},
  volume    = {99},
  journal   = {Phys. Rev. A},
  publisher = {APS},
  year      = {2019},
}

@Article{zhang2020subradiant,
  author    = {Zhang, Yu-Xiang and M{\o}lmer, Klaus},
  title     = {Subradiant emission from regular atomic arrays: universal scaling of decay rates from the generalized Bloch theorem},
  doi       = {https://doi.org/10.1103/PhysRevLett.125.253601},
  number    = {25},
  pages     = {253601},
  volume    = {125},
  journal   = {Phys. Rev. Lett.},
  publisher = {APS},
  year      = {2020},
}

@Article{eltohfa2025effects,
  author    = {Eltohfa, Mohamed and Robicheaux, Francis},
  title     = {Effects of finite trapping on the decay, recoil, and decoherence of dark states of quantum emitter arrays},
  doi       = {https://doi.org/10.1103/qv4c-s9xw},
  number    = {2},
  pages     = {023112},
  volume    = {112},
  journal   = {Phys. Rev. A},
  publisher = {APS},
  year      = {2025},
}

@Article{shi2025chiral,
  author    = {Shi, Jiaming and Poddubny, Alexander N},
  title     = {Chiral Dissociation of Bound Photon Pairs for a Non-Hermitian Skin Effect},
  doi       = {https://doi.org/10.1103/q6wr-2rt9},
  number    = {23},
  pages     = {233602},
  volume    = {134},
  journal   = {Phys. Rev. Lett.},
  publisher = {APS},
  year      = {2025},
}

@Article{marques2021bound,
  author    = {Marques, Y and Shelykh, IA and Iorsh, IV},
  title     = {Bound photonic pairs in 2D waveguide quantum electrodynamics},
  doi       = {10.1103/physrevlett.127.273602},
  number    = {27},
  pages     = {273602},
  volume    = {127},
  journal   = {Phys. Rev. Lett.},
  publisher = {APS},
  year      = {2021},
}

@Article{calajo2022emergence,
  author    = {Calajo, Giuseppe and Chang, Darrick E},
  title     = {Emergence of solitons from many-body photon bound states in quantum nonlinear media},
  doi       = {10.1103/physrevresearch.4.023026},
  number    = {2},
  pages     = {023026},
  volume    = {4},
  journal   = {Phys. Rev. Res.},
  publisher = {APS},
  year      = {2022},
}

@Article{tian2024power,
  author    = {Tian, Guoqing and Wu, Ying and L{\"u}, Xin-You},
  title     = {Power-law-exponential interaction induced quantum spiral phases},
  doi       = {10.1103/physrevresearch.6.033290},
  number    = {3},
  pages     = {033290},
  volume    = {6},
  journal   = {Phys. Rev. Res.},
  publisher = {APS},
  year      = {2024},
}

@Article{bjorn2022passive,
  author    = {Schrinski, Bj\"orn and Lamaison, Miren and S\o{}rensen, Anders S.},
  title     = {Passive Quantum Phase Gate for Photons Based on Three Level Emitters},
  doi       = {10.1103/PhysRevLett.129.130502},
  issue     = {13},
  pages     = {130502},
  url       = {https://link.aps.org/doi/10.1103/PhysRevLett.129.130502},
  volume    = {129},
  journal   = {Phys. Rev. Lett.},
  month     = {Sep},
  numpages  = {7},
  publisher = {American Physical Society},
  year      = {2022},
}

@Article{bello2019unconventional,
  author    = {Bello, Miguel and Platero, Gloria and Cirac, Juan Ignacio and Gonz{\'a}lez-Tudela, Alejandro},
  title     = {Unconventional quantum optics in topological waveguide QED},
  doi       = {10.1126/sciadv.aaw0297},
  number    = {7},
  pages     = {eaaw0297},
  volume    = {5},
  journal   = {Sci. Adv.},
  publisher = {American Association for the Advancement of Science},
  year      = {2019},
}

@Article{rusconi2021exploiting,
  author    = {Rusconi, Cosimo C and Shi, Tao and Cirac, J Ignacio},
  title     = {Exploiting the photonic nonlinearity of free-space subwavelength arrays of atoms},
  doi       = {https://doi.org/10.1103/PhysRevA.104.033718},
  number    = {3},
  pages     = {033718},
  volume    = {104},
  journal   = {Phys. Rev. A},
  publisher = {APS},
  year      = {2021},
}

@Article{castells2025cavity,
  author    = {Castells-Graells, David and Cirac, J Ignacio and Wild, Dominik S},
  title     = {Cavity quantum electrodynamics with atom arrays in free space},
  doi       = {https://doi.org/10.1103/PhysRevA.111.053712},
  number    = {5},
  pages     = {053712},
  volume    = {111},
  journal   = {Phys. Rev. A},
  publisher = {APS},
  year      = {2025},
}

@Article{needham2019subradiance,
  author    = {Needham, Jemma A and Lesanovsky, Igor and Olmos, Beatriz},
  title     = {Subradiance-protected excitation transport},
  doi       = {10.1088/1367-2630/ab31e8},
  number    = {7},
  pages     = {073061},
  volume    = {21},
  journal   = {New J. Phys.},
  publisher = {IOP Publishing},
  year      = {2019},
}

@Article{mitsch2014quantum,
  author    = {Mitsch, Rudolf and Sayrin, Cl{\'e}ment and Albrecht, Bernhard and Schneeweiss, Philipp and Rauschenbeutel, Arno},
  title     = {Quantum state-controlled directional spontaneous emission of photons into a nanophotonic waveguide},
  doi       = {10.1038/ncomms6713},
  number    = {1},
  pages     = {5713},
  volume    = {5},
  journal   = {Nat. Commun.},
  publisher = {Nature Publishing Group UK London},
  year      = {2014},
}

@article{poddubny2026bound,
  title={Bound, antibound and resonance two-photon states in chiral waveguide QED},
  author={Poddubny, Alexander},
  journal={arXiv:2604.20602},
  year={2026},
  url={https://doi.org/10.48550/arXiv.2604.20602}
}

@article{wang2022chiral,
  title={Chiral quantum network with giant atoms},
  author={Wang, Xin and Li, Hong-Rong},
  journal={Quantum Sci. Technol.},
  volume={7},
  number={3},
  pages={035007},
  year={2022},
  publisher={IOP Publishing},
  doi={10.1088/2058-9565/ac6a04}
}

@article{zhang2021charge,
  title={Charge-noise insensitive chiral photonic interface for waveguide circuit QED},
  author={Zhang, Yu-Xiang and i Carceller, Carles R and Kjaergaard, Morten and S{\o}rensen, Anders S},
  journal={Phys. Rev. Lett.},
  volume={127},
  number={23},
  pages={233601},
  year={2021},
  publisher={APS},
  doi={https://doi.org/10.1103/PhysRevLett.127.233601}
}

@article{fedorovich2022chirality,
  title={Chirality-driven delocalization in disordered waveguide-coupled quantum arrays},
  author={Fedorovich, Gleb and Kornovan, Danil and Poddubny, Alexander and Petrov, Mihail},
  journal={Phys. Rev. A},
  volume={106},
  number={4},
  pages={043723},
  year={2022},
  publisher={APS},
  doi={https://doi.org/10.1103/PhysRevA.106.043723}
}

@article{zhong2020photon,
  title={Photon-mediated localization in two-level qubit arrays},
  author={Zhong, Janet and Olekhno, Nikita A and Ke, Yongguan and Poshakinskiy, Alexander V and Lee, Chaohong and Kivshar, Yuri S and Poddubny, Alexander N},
  journal={Phys. Rev. Lett.},
  volume={124},
  number={9},
  pages={093604},
  year={2020},
  publisher={APS},
  doi={https://doi.org/10.1103/PhysRevLett.124.093604}
}

@article{le2005nanofiber,
  title={Nanofiber-mediated radiative transfer between two distant atoms},
  author={Le Kien, Fam and Gupta, S Dutta and Nayak, KP and Hakuta, K},
  journal={Phys. Rev. A},
  volume={72},
  number={6},
  pages={063815},
  year={2005},
  publisher={APS},
  doi={https://doi.org/10.1103/PhysRevA.72.063815}
}

@article{tong2004single,
  title={Single-mode guiding properties of subwavelength-diameter silica and silicon wire waveguides},
  author={Tong, Limin and Lou, Jingyi and Mazur, Eric},
  journal={Opt. Express},
  volume={12},
  number={6},
  pages={1025--1035},
  year={2004},
  publisher={Optical Society of America},
  doi={https://doi.org/10.1364/OPEX.12.001025}
}

@incollection{solano2017optical,
  title={Optical nanofibers: a new platform for quantum optics},
  author={Solano, Pablo and Grover, Jeffrey A and Hoffman, Jonathan E and Ravets, Sylvain and Fatemi, Fredrik K and Orozco, Luis A and Rolston, Steven L},
  booktitle={Advances In Atomic, Molecular, and Optical Physics},
  volume={66},
  pages={439--505},
  year={2017},
  publisher={Elsevier}
}

@article{svendsen2023modified,
  title={Modified dipole-dipole interactions in the presence of a nanophotonic waveguide},
  author={Svendsen, Mathias BM and Olmos, Beatriz},
  journal={Quantum},
  volume={7},
  pages={1091},
  year={2023},
  publisher={Verein zur F{\"o}rderung des Open Access Publizierens in den Quantenwissenschaften},
  doi={https://doi.org/10.22331/q-2023-08-22-1091}
}

\clearpage
\onecolumngrid
\begin{center}
    \textbf{\large Supplemental Material for\\[.2cm]``Most Subradiant Bound Photon Pairs from Chirality-Mediated Dispersion Softening''}\\[.4cm]
    Kailin Tan, Xuanbing Jiang, Dong Wang, and Saijun Wu\\
    \textit{Department of Physics, State Key Laboratory of Surface Physics and Key Laboratory of Micro\\ and Nano Photonic Structures (Ministry of Education), Fudan University, Shanghai 200433, China.}
\end{center}

\maketitle
\setcounter{equation}{0}
\setcounter{section}{0}
\setcounter{figure}{0}
\setcounter{table}{0}
\setcounter{page}{1}
\renewcommand{\theequation}{S\arabic{equation}}
\renewcommand{\thesection}{ \Roman{section}}

\renewcommand{\thefigure}{S\arabic{figure}}
\renewcommand{\thetable}{\arabic{table}}
\renewcommand{\tablename}{Supplementary Table}

\renewcommand{\bibnumfmt}[1]{[S#1]}
\renewcommand{\citenumfont}[1]{#1}
\makeatletter

\twocolumngrid

This Supplemental Material is structured as follows. We provide a general derivation of Eq.~\eqref{eq:Heff} in the main text in Sec.~\hyperref[sec:Heff]{S-I}. The derivation does not depend on the specific dispersion relation of the photonic reservoir. Then, we derive analytically the decay rates of subradiant FSs in Sec.~\hyperref[sec:FSproof]{S-II}, and detail the derivation of Eq.~\eqref{eq:BSscaling} in the main text for BSs in Sec.~\hyperref[sec:BSproof]{S-III}. Next, we characterize numerically the two-excitation eigenstates, especially the subradiant BSs in Sec.~\hyperref[sec:Character]{S-IV}. To further elucidate the subradiance mechanism, we construct and solve a supplemental toy model in Sec.~\hyperref[sec:toy]{S-V}. Finally, the details regarding the simulations of the nanofiber interface is given by Sec.~\hyperref[sec:Nanofiber]{S-VI}. 

We set $\hbar=1$ throughout this work.

\section{S-I. General derivation of $H_\mathrm{eff}$ \label{sec:Heff}}
In this section, we present a rigorous derivation of the effective Hamiltonian for a two-level atom array coupled chirally to a one-dimensional photonic reservoir. We assume the dispersion relation of the waveguide is given by a general form $\Omega_k$. Here, we require $\Omega_k$ to be monotonically increasing in the range $k \in [0, \Lambda]$, where $\Lambda$ is the momentum cutoff due to the waveguide geometry. Meanwhile, the Taylor expansion of $\Omega_k$ for $k > 0$ must contain only odd-power terms. Typical examples include $\Omega_k = v |k|$ for linear reservoirs and $\Omega_k = 2v/\sqrt{\delta_x} |\sin(k\delta_x/2)|$ for superconducting circuits~\cite{zhang2021charge}. The Hamiltonian in the interaction picture is $H_{\mathrm{int}}(t)=H_{\mathrm{int}}^{\mathrm{RW}}(t)+H_{\mathrm{int}}^{\mathrm{CRW}}(t)$, with the rotating-wave (RW) term
\begin{widetext}
\begin{equation}
    H_{\mathrm{int}}^{\mathrm{RW}}(t) =  \sum_{j=1}^{N} \int_{0}^{\Lambda} \frac{dk}{2\pi} \left[ \left( g_k e^{ikx_j} a_k  + h_{-k} e^{-ikx_j} a_{-k} \right) e^{i(\omega_0-\Omega_k)t} \sigma^{\dagger}_{j} + \text{h.c.} \right]
\label{eq:Hint}
\end{equation}
and the counter-rotating-wave (CRW) term
\begin{equation}
    H_{\mathrm{int}}^{\mathrm{CRW}}(t) = \sum_{j=1}^N \int_0^\Lambda \frac{dk}{2\pi} \left[ \left( g_k e^{ikx_j} a_k  + h_{-k} e^{-ikx_j} a_{-k} \right) e^{-i(\omega_0+\Omega_k)t} \sigma_j + \text{h.c.} \right].
\end{equation}
Here, $a_k$ is the waveguide photon operator and $\omega_0=\Omega_{k_0}$ is the atomic resonant frequency; $g_k$ and $h_{-k}$ are the coupling strengths between the atom and the right-/left-propagating photon, respectively, which have $|g_k|,|h_{-k}|\propto \sqrt{\Omega_k}$. In our derivation, $g_k$ and $h_{-k}$ are imposed to be real numbers. We start from the general form of the master equation under Born-Markov approximation
\begin{equation}
    \begin{split}
        \frac{\partial \rho}{\partial t} &=  -\int_{0}^{\infty }\Tr_R \left[ H_\mathrm{int}(t),[H_\mathrm{int}(t-\tau),\rho(t)\otimes \ketbra{V}] \right]     \\
    & = -\int_{0}^{\infty }\Tr_R \left[ H_\mathrm{int}(t)H_\mathrm{int}(t-\tau)\rho(t)\otimes \ketbra{V} \right]  + \dots,
    \end{split}
\label{eq:MasEqu}
\end{equation}
with $\rho$ the density operator of the atoms and $\ket{V}$ the vacuum state of waveguide photons. $\Tr_R$ represents the trace over the photonic reservoir. We treat the RW and the CRW contributions separately, i.e., $\partial_t\rho=(\partial_t\rho)_\mathrm{RW}+(\partial_t\rho)_\mathrm{CRW}$, with their cross terms omitted under the secular approximation. For the RW term, we get $(\partial_t\rho)_\mathrm{RW} = (\mathcal{L}_g^\mathrm{RW} + \mathcal{L}_h^\mathrm{RW})\rho + \dots$, where the Liouvillian superoperator $\mathcal{L}_g^\mathrm{RW}$ related to $g_k$ reads explicitly
\begin{equation}
    \begin{split}
        \mathcal{L}_g^\mathrm{RW} & =  -\sum_{j<l}^N \int_{0}^{\infty} d\tau     \int_0^{\Lambda} \frac{dk}{2\pi}  g_k^2 e^{i\tau(\omega_0-\Omega_k)}  
    \left( e^{ik(x_j-x_l)}\sigma_j^\dagger \sigma_l + e^{-ik(x_j-x_l)}\sigma_l^\dagger \sigma_j \right) \\
    &\quad - \sum_{j=1}^N \int_{0}^{\infty} d\tau\int_0^{\Lambda} \frac{dk}{2\pi }   g_k^2 e^{i\tau(\omega_0-\Omega_k)} \sigma_j^\dagger\sigma_j.
    \end{split}
\end{equation}
Using the Heitler identity $\int_0^{\infty} d \tau e^{i\tau\left(\omega_{0}-\Omega_k\right) }=\pi \delta\left(\omega_{0}-\Omega_k\right)+i \mathcal{P} \frac{1}{\omega_{0}-\Omega_k}$ (with $\mathcal{P}$ the Cauchy principal value) and the substitution $\delta (\omega_0-\Omega_k) =  \frac{1}{v_g} \left[ \delta(k-k_0) + \delta (k+k_0) \right]$ with the group velocity $v_g = \partial_k\Omega_k|_{k=k_0}$ is the group velocity, we have $\mathcal{L}_g^\mathrm{RW} = \mathcal{L}_g^{\mathrm{RW},0} + \mathcal{L}_g^{\mathrm{RW},1} + \mathcal{L}_g^{\mathrm{RW},2}$. Here, 
\begin{subequations}
\begin{align}
    \mathcal{L}_g^{\mathrm{RW},0} & = -\frac{1}{2v_g} \sum_{j=1}^{N} g_{k_0}^2 \sigma_j^\dagger \sigma_j , \\
    \mathcal{L}_g^{\mathrm{RW},1} & = -\frac{1}{2v_g} \sum_{j<l}^N g_{k_0}^2 \left[ \sigma_j^\dagger\sigma_l e^{ik_0(x_j-x_l)} + \text{h.c.} \right],\\
    \mathcal{L}_g^{\mathrm{RW},2} & = \mathcal{L}_g^{\mathrm{RW},2a} + \mathcal{L}_g^{\mathrm{RW},2b}
         = -i \sum_{j<l}^N \mathcal{P} \int_0^{\Lambda} \frac{dk}{2\pi } g_{k}^2 \left[ \sigma_j^\dagger\sigma_l \frac{e^{ik(x_j-x_l)}}{\omega_0-\Omega_k} + \text{h.c.} \right].
\end{align}
\end{subequations}
Note that we have renormalized the divergent self-energy term. Performing contour integrals to compute $\mathcal{L}_g^{\mathrm{RW},2a}$ and $\mathcal{L}_g^{\mathrm{RW},2b}$, we obtain
\begin{subequations}
\begin{align}
    \mathcal{L}_g^{\mathrm{RW},2a} &= -i \sum_{j<l}^N \sigma_j^\dagger\sigma_l \left[ \frac{i}{2v_g} g_{k_0}^2 e^{ik_0(x_j-x_l)} 
    + \int_0^{-\Lambda} \frac{i du}{2\pi } g_{iu}^2 \frac{e^{-u(x_j-x_l)}}{\omega_0-\Omega_{iu}}\right],      \\
    \mathcal{L}_g^{\mathrm{RW},2b} &= -i \sum_{j<l}^N \sigma_l^\dagger\sigma_j \left[ \frac{-i}{2v_g} g_{k_0}^2 e^{-ik_0(x_j-x_l)} 
    + \int_0^{\Lambda} \frac{i du}{2\pi }  g_{iu}^2 \frac{e^{u(x_j-x_l)}}{\omega_0-\Omega_{iu}}\right].
\end{align}
\end{subequations}
Notably, the analytic continuation here is performed for $k>0$, rather than on the entire real axis, since the absolute value function $|k|$ is non-analytic when we consider $\Omega_{-k}=\Omega_{|k|}$ for $k<0$. Meanwhile, as $\Omega_k$ contains only odd-power terms in its Taylor expansion for $k>0$, the analytic continuation ensures $\Omega_{-iu} = -\Omega_{iu}$. Moreover, the coupling satisfies $g_{-iu}^2 = -g_{iu}^2$ for we have $g_k^2\propto \Omega_k$. Summing the above terms yields
\begin{equation}
    \begin{split}
        \mathcal{L}_g^\mathrm{RW}&=-\frac{g_{k_0}^2}{2v_g} \sum\limits_{j=1}^N \sigma_j^\dagger\sigma_j - \frac{g_{k_0}^2}{v_g} \sum\limits_{j<l}^N e^{ik_0(x_l-x_j)}\sigma_j^\dagger\sigma_l\\
        &\quad -i \sum_{j<l}^N\left[\sigma_j^\dagger\sigma_l \int_0^{-\Lambda} \frac{i du}{2\pi } g_{iu}^2 \frac{e^{-u(x_j-x_l)}}{\omega_0-\Omega_{iu}} + \sigma_l^\dagger\sigma_j  \int_0^{\Lambda} \frac{i du}{2\pi } g_{iu}^2 \frac{e^{u(x_j-x_l)}}{\omega_0-\Omega_{iu}}\right].
    \end{split}
\end{equation}
On the other hand, the CRW contribution is $(\partial_t\rho)_\mathrm{CRW} = (\mathcal{L}_g^\mathrm{CRW} + \mathcal{L}_h^\mathrm{CRW})\rho + \dots$ After some algebra, we get the Liouvillian
\begin{equation}
    \mathcal{L}_g^\mathrm{CRW}= i \sum_{j<l}^N\left[\sigma_j\sigma_l^\dagger \int_0^{-\Lambda} \frac{i du}{2\pi } g_{iu}^2 \frac{e^{-u(x_j-x_l)}}{\omega_0+\Omega_{iu}} + \sigma_l\sigma_j^\dagger  \int_0^{\Lambda} \frac{i du}{2\pi } g_{iu}^2 \frac{e^{u(x_j-x_l)}}{\omega_0+\Omega_{iu}}\right].
\end{equation}
Using the relations $\Omega_{-iu} = -\Omega_{iu}$ and $g_{-iu}^2 = -g_{iu}^2$, we find
\begin{equation}
    \mathcal{L}_g=\mathcal{L}_g^\mathrm{RW}+\mathcal{L}_g^\mathrm{CRW}=-\frac{g_{k_0}^2}{2v_g} \sum\limits_{j=1}^N \sigma_j^\dagger\sigma_j - \frac{g_{k_0}^2}{v_g} \sum\limits_{j<l}^N e^{ik_0(x_l-x_j)}\sigma_j^\dagger\sigma_l.
\end{equation}
Crucially, the integral terms are exactly canceled. Similarly, the Liouvillian $\mathcal{L}_h=\mathcal{L}_h^\mathrm{RW}+\mathcal{L}_h^\mathrm{CRW}$ related to $h_{-k}$ reads
\begin{equation}
    \mathcal{L}_h=-\frac{h_{-k_0}^2}{2v_g} \sum\limits_{j=1}^N \sigma_j^\dagger\sigma_j - \frac{h_{-k_0}^2}{v_g} \sum\limits_{j<l}^N e^{ik_0(x_l-x_j)}\sigma_j^\dagger\sigma_l.
\end{equation}
Comparing the equation $\partial_t\rho = (\mathcal{L}_g + \mathcal{L}_h)\rho + \dots$ here with ${\partial_t}\rho= -i H_{\mathrm{eff}} \rho+\dots$, we extract the effective Hamiltonian, i.e., Eq.~\eqref{eq:Heff} in the main text, where the decay rates are $\Gamma_R = 2g_{k_0}^2/v_g$ and $\Gamma_L = 2h_{-k_0}^2/v_g$.
\end{widetext}

\section{S-II. Subradiant Free-Fermion states in chiral waveguide QED\label{sec:FSproof}}
In this section, we present the analytical derivation regarding the subradiant FSs in chiral waveguide QED. We solve the single-excitation subradiant states, and then construct the FSs using these single-excitation states.

\subsection{A. subradiant single-excitation states}
\begin{figure}[htbp]
    \includegraphics[scale=0.59]{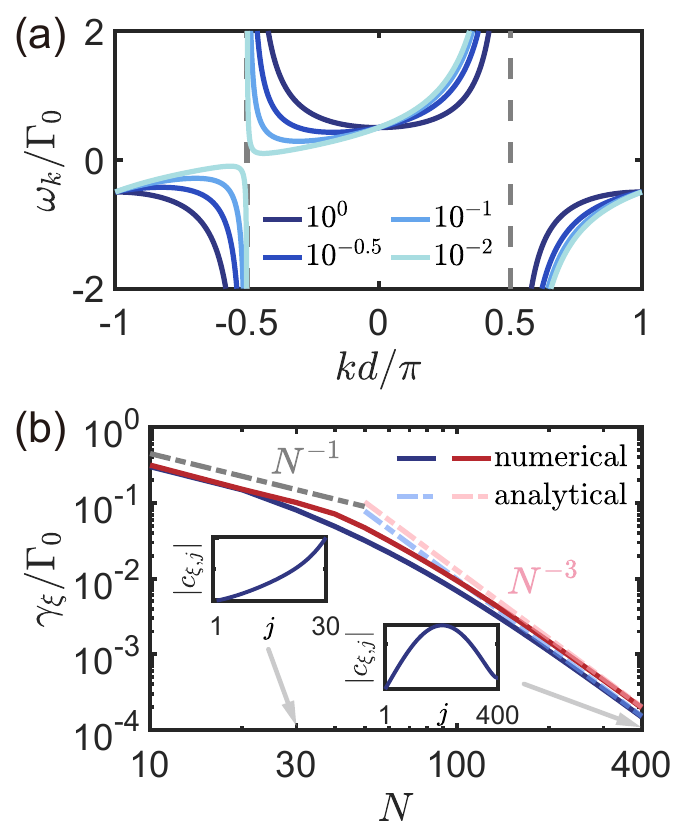}
    \centering
    \caption{(a) Single-excitation dispersion relation of chiral waveguide QED, with $d=0.25\lambda_0$ and $\eta_{LR}=1,10^{-0.5},10^{-1}$ and $10^{-2}$, respectively. The dashed lines are the light cones where $\omega_k$ diverges. (b) Numerical (solid lines) and analytical (dashed dotted lines) decay rate scaling of the single-excitation subradiant states, for $k_\mathrm{ex}=k_\mathrm{ex}^{(-)}$ (blue) and $k_\mathrm{ex}=k_\mathrm{ex}^{(+)}$ (red), respectively. Insets: probability amplitude distribution of the subradiant states near $k_\mathrm{ex}^{(-)}$, with $N=30$ (left) and $N=400$ (right). Here, we choose the spacing $d=0.12\lambda_0$.} 
    \label{fgS:single} 
\end{figure}

For an infinite array, the effective Hamiltonian has the Bloch eigenstate $\ket*{k}=\sum_{j=-\infty}^\infty e^{ikx_j}\sigma_j^\dagger\ket*{G}$, with $k\in[-\pi/d,\pi/d]$ the pseudo wavevector and $\ket*{G}=\bigotimes_{j}\ket*{g_j}$ the collective ground state. Solving the eigenequation $H_\text{eff}\ket*{k}=\omega_k\ket*{k}$ yields
\begin{equation}
    \omega_k=\frac{\Gamma_L}{4}\cot\frac{(k_0+k)d}{2}+\frac{\Gamma_R}{4}\cot\frac{(k_0-k)d}{2}.
    \label{eq:omegak}
\end{equation}
The single-excitation polaritonic dispersion $\omega_k$ for different chirality is shown in Fig.~\hyperref[fgS:single]{S1(a)}. For $\eta_{LR}\neq1$, the extremum points are given by $\partial_k\omega_k|_{k_\mathrm{ex}}=0$, which reads explicitly:
\begin{equation}
    k_\text{ex}^{(\pm)}=k_0-\frac{2}{d}\cot^{-1}\left(\frac{\cos k_0d\pm\sqrt{\eta_{LR}}}{\sin k_0d}\right).
    \label{eq:kex}
\end{equation}

For a finite array, we can construct the timed-Dicke state $\ket*{\beta_k}=N^{-1/2}\sum_{j=1}^N e^{ikx_j}\sigma_j^\dagger\ket*{G}$ and compute directly
\begin{equation}    
    H_\text{eff}\ket*{\beta_k}=\omega_k\ket*{\beta_k}-\frac{i}{2}(\Gamma_R v_k\ket*{\beta_{k_0}}-\Gamma_L w_k\ket*{\beta_{-k_0}}). 
\label{eq:singleDirect}
\end{equation}
Here, the tails have $v_k=\frac{e^{i(k-k_0)d}}{1-e^{i(k-k_0)d}}$ and $w_k=\frac{e^{i(k+k_0)Nd}}{e^{-i(k+k_0)d}-1}$. For an arbitrary $\omega_k$ allowed by the spectrum, there exist two wavevectors $k_\pm$ that satisfy $\omega_{k_+}=\omega_{k_-}$ and thus have the same diagonal element in Eq.~\eqref{eq:singleDirect}. Hence, we expect the linear combination of $\ket*{\beta_{k_\pm}}$ to be the eigenstate with tails canceled, which yields the equation
\begin{equation}
    v_{k_+}w_{k_-}=v_{k_-}w_{k_+}.
    \label{eq:GBZ}
\end{equation}
For subradiant states, we denote $k_\pm=k_\text{ex}\pm\delta k_\xi$ with $1/N\ll\sqrt{\eta_{LR}}$ and find
\begin{equation}
    \delta k_\xi\approx \frac{\xi\pi}{Nd}\left[1-\frac{i}{N}\frac{\sin k_0d}{2\sin\frac{(k_0+k_\mathrm{ex})d}{2}\sin\frac{(k_0-k_\mathrm{ex})d}{2}}\right],
    \label{eq:deltaxi}
\end{equation}
with $\xi=1,2,\dots, \xi\ll N$. Substituting $\delta k_\xi$ into the dispersion relation $\omega_\xi=\omega_{k_\text{ex}\pm\delta k_\xi}$ yields the decay rate $\gamma_\xi=-2\text{Im}\omega_\xi$, which is given by Eq.~\eqref{eq:singleScaling} in the main text. Figure~\hyperref[fgS:single]{S1(b)} shows the comparison between the analytical decay rates and the corresponding numerical results. The analytical decay rates coincides with the numerical ones for a sufficiently large $N$. It can also be observed that $\gamma_\xi$ scales as $N^{-1}$ for small $N$, and the subradiant eigenstate $\ket*{\psi_\xi}=\sum_{j=1}^N c_{\xi,j}\sigma_j^\dagger\ket*{G}$ are localized at the edge, which were first reported in Ref.~\cite{fedorovich2022chirality}.

\subsection{B. Free-Fermion states}
With the analytical understanding of the single-excitation subradiant states, we can therefore construct the FSs. Here we present two approaches. The first one follows Refs.~\cite{zhang2019theory,schrinski2022polariton} that maps $H_\mathrm{eff}$ to the Tonks-Girardeau gas. Using the Holstein-Primakoff transformation (expanded to the second order) $\sigma_j\approx (1-b_j^\dagger b_j/2)b_j$, we find that the imaginary part of the Hamiltonian $H_\mathrm{eff}^\mathrm{Im}=-\text{Im}H_\mathrm{eff}$ governing the collective emission reads $H_\text{eff}^\text{Im}=H_\text{sup}+V_\text{sca}+V_\text{sca}^\dagger$, with
\begin{subequations}
    \begin{align}
        H_\text{sup}&=\frac{N}{4}\left( \Gamma_L b_{-k_0}^\dagger b_{-k_0}+ \Gamma_R b_{k_0}^\dagger b_{k_0}\right),\\
        V_\text{sca}&=-\frac{1}{8}\sum\limits_{k,q}\left(\Gamma_L b_{-k_0}^\dagger b_{k+q+k_0}^\dagger b_k b_q \notag \right.\\
        & \left. \qquad\qquad\qquad+\Gamma_R b_{k_0}^\dagger b_{k+q-k_0}^\dagger b_k b_q \right).
    \end{align}
    \label{eq:HsupVsca}
\end{subequations}
Here, the Fourier transform of the Bosonic operator is $b_k^\dagger=N^{-1/2}\sum_{j=1}^N e^{ikx_j} b_j^\dagger$. After eliminating the superradiant states, we obtain the effective theory $V_\mathrm{sub}$ for the subradiant states, which is the same as Eq.~(8) of Ref.~\cite{zhang2019theory}. This indicates that chiral emission does not affect the scattering dynamics in the subradiant subspace. Hence, $V_\mathrm{sub}$ can be likewise mapped to the Lieb-Liniger model, where the free-Fermion eigenstates occur in the Tonks-Girardeau limit.

The other approach is based on the Jordan-Wigner transformation $\sigma_j^\dagger=e^{i\pi\sum_{l=1}^{j-1}f_l^\dagger f_l}f_j^\dagger$. Since $\omega_k$ is quadratic near the extremum points, one can utilize a simpler tight-binding Hamiltonian $\mathbf{H}_1$ to approximate $H_\mathrm{eff}$ and find that $\mathbf{H}_1$ and $H_\mathrm{eff}$ share the same free-Fermion multi-excitation eigenstates in the leading order, as detailed in Ref.~\cite{zhang2022free}. 

\begin{figure}[htbp]
    \includegraphics[scale=0.585]{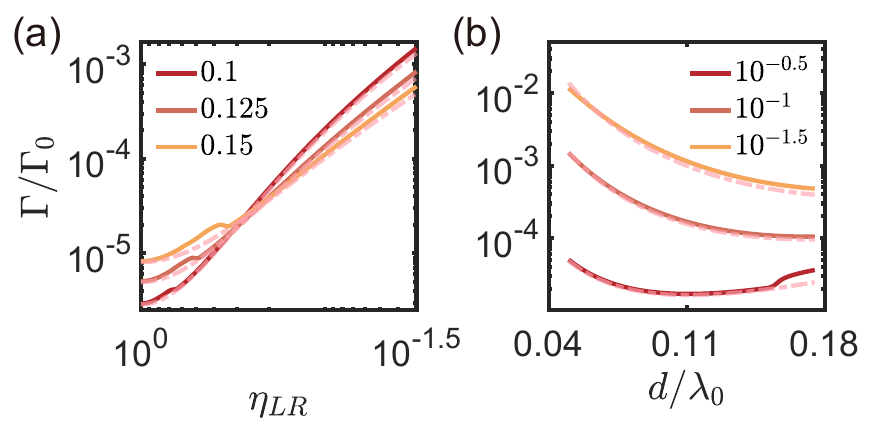}
    \centering
    \caption{Minimal FS decay rates [for ${\bm{\xi}}=(1,2)$] as a function of $\eta_{LR}$ (a) and $d$ (b) for $N=100$. Solid and dashed dotted curves represent the numerical and analytical results, respectively. Legends: the values of $d/\lambda_0$ (a) and $\eta_{LR}$ (b).} 
    \label{fgS:FSdeta} 
\end{figure}

The FS wavefunction reads
\begin{equation}
    \ket*{\Phi_{\bm{\xi}}}=\prod\limits_{s=1}^{n_e} f_{\xi_s}^\dagger\ket*{G}+\mathcal{O}\left(\frac{1}{N}\right).
    \label{eq:FS}
\end{equation}
Here, ${\bm{\xi}}=(\xi_1,\dots,\xi_{n_e})$ are the excitation indices with total excitation number $n_e$ and $f_{\xi}^\dagger=\sum_{j=1}^N c_{\xi,j}f_j^\dagger$ is the Fermionic operator. The corresponding eigen decay rate is $\Gamma_{\bm{\xi}}=\sum_{s=1}^{n_e}\gamma_{\xi_s}$, i.e., the expression given in the main text. We have shown in Fig.~\hyperref[fg:BSeAlpha2]{2(c)} of the main text the FS decay rate scaling, where the analytical predictions are consistent with the numerical results. In addition, we also compute the dependence of the minimal FS decay rates on $\eta_{LR}$ and $d$, as displayed in Fig.~\ref{fgS:FSdeta}.

\section{S-III. Subradiant bound state decay rates from quadratic band \label{sec:BSproof}}
Here, we provide the detailed proof of Eq.~\eqref{eq:BSscaling} in the main text. As we find $H_\mathrm{eff}\ket*{K,\Delta}=H^K\ket*{K,\Delta}+\ket*{T_{K,\Delta}}$ with $H^K=\sum_{\Delta,\Delta'=d}^{(N-1)d}\mathcal{H}_{\Delta\Delta'}^K\ketbra*{K,\Delta}{K,\Delta'}$, we can first diagonalize $H^K$ and then work on the cancellation of the tails.

\subsection{A. Bound eigenstates of $H^K$}

We first study the bound eigenstates of the effective single-particle Hamiltonian $H^K$. For an infinite array, the exact bound solution 
\begin{equation}
    \ket*{\psi_K}_\infty=\sum\limits_{\Delta/d=1}^\infty [A_K(z_1^K)^{\Delta/d}+B_K(z_2^K)^{\Delta/d}]\ket*{K,\Delta}_\infty
\end{equation}
can be obtained from the equations~\cite{bakkensen2021photonic}
\begin{subequations}
    \begin{align}
        F(K,z_1^K)&=F(K,z_2^K),\\
        g_+(z_1^K)g_-(z_2^K)&=g_-(z_1^K)g_+(z_2^K),
    \end{align}
\end{subequations} 
with
\begin{equation}
    \begin{split}
        F(K,z)&=\frac{\Gamma_L z\sin(k_0+\frac{K}{2})d}{1+z^2-2z\cos(k_0+\frac{K}{2})d}\\
        &\quad +\frac{\Gamma_R z\sin(k_0-\frac{K}{2})d}{1+z^2-2z\cos(k_0-\frac{K}{2})d},
    \end{split}
\end{equation}
and
\begin{subequations}
    \begin{align}
        g_+(z)&=\Gamma_L z\frac{z-\cos(k_0+\frac{K}{2})d}{1+z^2-2z\cos(k_0+\frac{K}{2})d},\\
        g_-(z)&=\Gamma_R z\frac{z-\cos(k_0-\frac{K}{2})d}{1+z^2-2z\cos(k_0-\frac{K}{2})d}.
    \end{align}
\end{subequations}
We can therefore get $z_{1,2}^K$ with $|z_{1,2}^K|<1$ and find $A_K/B_K=-g_-(z_1^K)/g_+(z_1^{K})$. The corresponding eigenenergy is $\mathcal{E}_K=F(K,z_{1,2}^K)$. To determine $A_K$ and $B_K$, we further impose the normalization condition $\sum_{\Delta/d=1}^\infty |C_\Delta^K|^2=1$. Note that $C_\Delta^K$ can be purely real.

\begin{figure}[htbp]
    \includegraphics[scale=0.58]{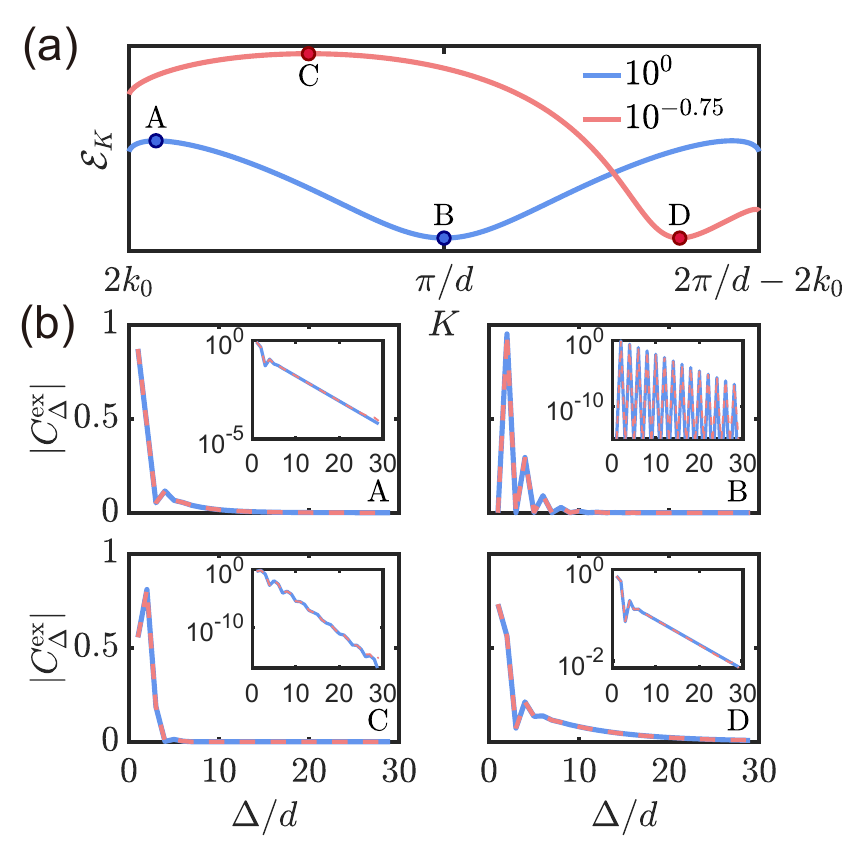}
    \centering
    \caption{(a) BS dispersion relations for $\eta_{LR}=1$ (blue) and $10^{-0.75}$ (red). (b) Bound eigenstates of $H^K$ at the extremum points A, B, C and D marked in (a). Solid (dashed) lines show the exact (diagonalized) solutions of an infinite array (a finite array with $N=30$). The insets are the same as the main plots but in log scale. Here, the spacing $d=0.15\lambda_0$ is used.} 
    \label{fgS:HKeig} 
\end{figure}

For a finite array with $N$ atoms, the eigen BS of $H^K$ can be approximated by $\braket*{K,\Delta}{\psi_K} \approx C_\Delta^K$. Physically, this arises from that $C_{(N-1)d}^K\sim\max\{\mathrm{Re}z_1^K,\mathrm{Re}z_2^K\}^N$ is exponentially small for $N\gg1$, so the scattering by the boundary is sufficiently weak. We show numerically in Fig.~\ref{fgS:HKeig} that the difference of $|\braket*{K,\Delta}{\psi_K}|$ at $K_\mathrm{ex}$ between an infinite array and a finite array is exponentially small. Therefore, we can safely omit this difference in our analytical treatment hereinafter. We denote
\begin{equation}
    \ket*{\psi_K}=\frac{1}{\sqrt{M_K}}\sum\limits_{\Delta/d=1}^{N-1}[A_K(z_1^K)^{\Delta/d}+B_K(z_2^K)^{\Delta/d}]\ket*{K,\Delta}.
\end{equation}
Here, the normalization factor $M_K$ has
\begin{equation}
    M_K=N-S_K,\quad S_K=\sum\limits_{\Delta/d=1}^{N-1} |C_\Delta^K|^2\Delta/d\sim\mathcal{O}(1),
\end{equation}
which is derived using the fact that $\braket*{K,\Delta}{K',\Delta'}=(N-\Delta/d)\delta_{KK'}\delta_{\Delta\Delta'}$. We can hence write down the orthonormal relation
\begin{equation}
    \braket*{\psi_K}{\psi_{K'}}=\delta_{KK'}.
\end{equation}

It is sufficient to study the cases with $0<d<\lambda_0/4$ for $H^K$. For this to be true, first, we apply the transformation $k_0d\to\pi-k_0d$ and find that
\begin{equation}
    H^K \to -U(H^{-K})^* U^\dagger,\quad U=\sum\limits_{\Delta}(-1)^{\Delta/d}\ketbra*{K,\Delta}{K,\Delta}.
\end{equation}
Hence, the eigenvalues of $H^K$ transforms correspondingly:
\begin{subequations}
    \begin{align}
        \text{Re}\mathcal{E}_K&\to-\text{Re}\mathcal{E}_{-K},\\
        \text{Im}\mathcal{E}_K&\to\text{Im}\mathcal{E}_{-K}.
    \end{align}
\end{subequations}
Note that the energy spectrum inversion is first reported in Fig.~4(a) of Ref.~\cite{calajo2022emergence} in a non-chiral system. Second, performing the transformation $k_0d\to k_0d+\pi$ yields
\begin{equation}
    H^K \to U H^{K} U^\dagger.
\end{equation}
The dispersion relation is therefore unchanged. With these transformations, we can obtain $\mathcal{E}_K$ for arbitrary $d>0$ from the results for $0<d<\lambda_0/4$.

\subsection{B. Full expression of the tails}
The tail of Eq.~\eqref{eq:HeffonKDelta} can be written explicitly:
\begin{equation}
    \begin{split}
        \ket*{T_{K,\Delta}}&=i\frac{N\Gamma_L}{2}e^{i\left(k_0+\frac{K}{2}\right)\Delta}\sigma_{-k_0}^\dagger (\sigma_{K+k_0}^\dagger)_{R,\Delta}\ket*{G}\\
        &\quad+i\frac{N\Gamma_R}{2}e^{i\left(k_0-\frac{K}{2}\right)\Delta}\sigma_{k_0}^\dagger (\sigma_{K-k_0}^\dagger)_{L,\Delta}\ket*{G}.
    \end{split}
\end{equation}
Here, the subscripts ``$R,\Delta$'' and ``$L,\Delta$'' represent the summations
\begin{subequations}
    \begin{align}
        \left(\sum\limits_{j=1}^N \circ\right)_{R,\Delta}&=\sum\limits_{j=N-\Delta/d+1}^N\circ  , \label{eq:sumR}\\
        \left(\sum\limits_{j=1}^N \circ\right)_{L,\Delta}&=\sum\limits_{j=1}^{\Delta/d}\circ.\label{eq:sumL}
    \end{align}
\end{subequations}
Diagonalizing $H^K$ yields the tails of Eq.~\eqref{eq:HeffonpsiK}, which read
\begin{subequations}
    \begin{align}
        \ket*{T_R}&=\sum\limits_{\Delta/d=1}^{N-1} i C_\Delta^K \frac{N\Gamma_L}{2\sqrt{M_K}}e^{i\left(k_0+\frac{K}{2}\right)\Delta}\sigma_{-k_0}^\dagger (\sigma_{K+k_0}^\dagger)_{R,\Delta}\ket*{G},\\
        \ket*{T_L}&=\sum\limits_{\Delta/d=1}^{N-1} i C_\Delta^K \frac{N\Gamma_R}{2\sqrt{M_K}}e^{i\left(k_0-\frac{K}{2}\right)\Delta}\sigma_{k_0}^\dagger (\sigma_{K-k_0}^\dagger)_{L,\Delta}\ket*{G}.
    \end{align}
\end{subequations}
Exchanging the summation order with respect to $\Delta$ and $j$ [in Eqs.~\eqref{eq:sumR} and~\eqref{eq:sumL}], and employing the substitutions $j=N-m+1$ for $\ket*{T_R}$ and $j=m$ for $\ket*{T_L}$, respectively, we obtain
\begin{widetext}
\begin{subequations}
    \begin{align}
        \ket*{T_R}&=i\frac{\Gamma_L}{2}\left(1+\frac{S_K}{2N}\right)\sigma_{-k_0}^\dagger\sum\limits_{m=1}^{N-1} e^{i(K+k_0)(N-m+1)d}\sigma_{N-m+1}^\dagger \notag\\
        &\quad \times\left( A_K\frac{\left[z_1^Ke^{i(k_0+\frac{K}{2})d}\right]^m}{1-z_1^K e^{i(k_0+\frac{K}{2})d}}+B_K\frac{\left[z_2^K e^{i(k_0+\frac{K}{2})d}\right]^m}{1-z_2^K e^{i(k_0+\frac{K}{2})d}} \right)\ket*{G} + \mathcal{O}(N^{-2}),\\
        \ket*{T_L}&=i\frac{\Gamma_R}{2}\left(1+\frac{S_K}{2N}\right)\sigma_{k_0}^\dagger\sum\limits_{m=1}^{N-1} e^{i(K-k_0)md}\sigma_m^\dagger \notag \\
        &\quad \times \left( A_K\frac{\left[z_1^Ke^{i(k_0-\frac{K}{2})d}\right]^m}{1-z_1^K e^{i(k_0-\frac{K}{2})d}}+B_K\frac{\left[z_2^K e^{i(k_0-\frac{K}{2})d}\right]^m}{1-z_2^K e^{i(k_0-\frac{K}{2})d}} \right)\ket*{G} + \mathcal{O}(N^{-2}).
    \end{align}
\end{subequations}
These tail states are strongly localized at their corresponding edges and vanish for $N\to\infty$. 

\subsection{C. Expanding tails near $K_\mathrm{ex}$}
As mentioned in the main text, we expect subradiant BSs to occur at $K\approx K_\mathrm{ex}$. Here, we perform the Taylor expansions on $K=K_\mathrm{ex}+\delta K$ with $\delta K\sim \mathcal{O}(N^{-1})$. We denote $\delta K=1/(\varsigma Nd)$, where $\varsigma\sim\mathcal{O}(1)$ is yet undetermined. Consider the right-edge state as an example, which can be expanded as 
\begin{equation}
        \ket*{T_R}=i\frac{\Gamma_L}{2}\sigma_{-k_0}^\dagger\sum\limits_{m=1}^{N-1} e^{i\delta Kd(N-m+1+m\tilde{\mu}_{R,m}+\tilde{\nu}_{R,m})} e^{i(K_\mathrm{ex}+k_0)(N-m+1)d}R_m \sigma_{N-m+1}^\dagger\ket*{G} + \mathcal{O}(N^{-2}),
\end{equation}
with the factors
\begin{subequations}
    \begin{align}
        \tilde{\mu}_{R,m}&=\frac{-i/d}{A_\mathrm{ex}\left[1-z_2^\mathrm{ex}e^{i(k_0+\frac{K_\mathrm{ex}}{2})d}\right]\left[z_1^\mathrm{ex}e^{i(k_0+\frac{K_\mathrm{ex}}{2})d}\right]^m+B_\mathrm{ex}\left[1-z_1^\mathrm{ex}e^{i(k_0+\frac{K_\mathrm{ex}}{2})d}\right]\left[z_2^\mathrm{ex}e^{i(k_0+\frac{K_\mathrm{ex}}{2})d}\right]^m} \notag \\
        &\quad \times\left( A_\mathrm{ex}\left[1-z_2^\mathrm{ex}e^{i(k_0+\frac{K_\mathrm{ex}}{2})d}\right]\left[z_1^\mathrm{ex}e^{i(k_0+\frac{K_\mathrm{ex}}{2})d}\right]^m \frac{\partial_K z_1^\mathrm{ex}+iz_1^\mathrm{ex}d/2}{z_1^\mathrm{ex}} \right. \notag\\
        &\left. \qquad +  B_\mathrm{ex}\left[1-z_1^\mathrm{ex}e^{i(k_0+\frac{K_\mathrm{ex}}{2})d}\right]\left[z_2^\mathrm{ex}e^{i(k_0+\frac{K_\mathrm{ex}}{2})d}\right]^m \frac{\partial_K z_2^\mathrm{ex}+iz_2^\mathrm{ex}d/2}{z_2^\mathrm{ex}} \right),\\
        \tilde{\nu}_{R,m}&=-\frac{i}{2}\varsigma S_\mathrm{ex} + \frac{-i/d}{A_\mathrm{ex}\left[1-z_2^\mathrm{ex}e^{i(k_0+\frac{K_\mathrm{ex}}{2})d}\right]\left[z_1^\mathrm{ex}e^{i(k_0+\frac{K_\mathrm{ex}}{2})d}\right]^m+B_\mathrm{ex}\left[1-z_1^\mathrm{ex}e^{i(k_0+\frac{K_\mathrm{ex}}{2})d}\right]\left[z_2^\mathrm{ex}e^{i(k_0+\frac{K_\mathrm{ex}}{2})d}\right]^m} \notag \\
        &\quad \times\left( A_\mathrm{ex}\left[1-z_2^\mathrm{ex}e^{i(k_0+\frac{K_\mathrm{ex}}{2})d}\right]\left[z_1^\mathrm{ex}e^{i(k_0+\frac{K_\mathrm{ex}}{2})d}\right]^m \left[\frac{e^{i\left(k_0+\frac{K_\mathrm{ex}}{2}\right)d}(\partial_K z_1^\mathrm{ex}+iz_1^\mathrm{ex}d/2)}{1-z_1^\mathrm{ex}e^{i(k_0+\frac{K_\mathrm{ex}}{2})d}}+\frac{\partial_K A_\mathrm{ex}}{A_\mathrm{ex}}\right] \right. \notag\\
        &\left. \qquad +  B_\mathrm{ex}\left[1-z_1^\mathrm{ex}e^{i(k_0+\frac{K_\mathrm{ex}}{2})d}\right]\left[z_2^\mathrm{ex}e^{i(k_0+\frac{K_\mathrm{ex}}{2})d}\right]^m \left[\frac{e^{i\left(k_0+\frac{K_\mathrm{ex}}{2}\right)d}(\partial_K z_2^\mathrm{ex}+iz_2^\mathrm{ex}d/2)}{1-z_2^\mathrm{ex}e^{i(k_0+\frac{K_\mathrm{ex}}{2})d}}+\frac{\partial_K B_\mathrm{ex}}{B_\mathrm{ex}}\right] \right),  
    \end{align}
\end{subequations}
and the expression of $R_m$ here is given in the main text. We can show that both $\tilde{\mu}_{R,m}$ and $\tilde{\nu}_{R,m}$ are bounded complex sequences in $m$. This property will be used in Secs.~\hyperref[subsec:constructH]{S-III.D} and~\hyperref[subsec:DeltaH]{S-III.E} to ensure that the complex phase term $\delta Kd(-m+1+m\tilde{\mu}_{R,m}+\tilde{\nu}_{R,m})$ is of the order $\mathcal{O}(N^{-1})$ for a sufficiently large $N$ and $\forall m$. The proofs of the boundedness are given below.

\subsubsection{1. Boundedness of $\tilde{\mu}_{R,m}$}
Here we prove that $\tilde{\mu}_{R,m}$ is a bounded sequence. To this end, we shall show that $|\tilde{\mu}_{R,m}|$ is bounded. First, if $z_1^\mathrm{ex}=z_2^\mathrm{ex}$, the proof is trivial since $\tilde{\mu}_{R,m}$ is independent of $m$. Thus, we focus on the case where $z_1^\mathrm{ex}\neq z_2^\mathrm{ex}$. We find that
\begin{equation}
    \begin{split}
        |\tilde{\mu}_{R,m}|&\leq \frac{1}{d} \left( \frac{1}{\abs{\abs{\frac{B_\mathrm{ex}}{A_\mathrm{ex}}\frac{1-z_1^\mathrm{ex}e^{i(k_0+K_\mathrm{ex}/2)d}}{1-z_2^\mathrm{ex}e^{i(k_0+K_\mathrm{ex}/2)d}}}\abs{\frac{z_2^\mathrm{ex}}{z_1^\mathrm{ex}}}^m-1}} \abs{\frac{\partial_K z_1^\mathrm{ex}+iz_1^\mathrm{ex}d/2}{z_1^\mathrm{ex}}} \right.\\
        &\left.\quad + \frac{1}{\abs{\abs{\frac{A_\mathrm{ex}}{B_\mathrm{ex}}\frac{1-z_2^\mathrm{ex}e^{i(k_0+K_\mathrm{ex}/2)d}}{1-z_1^\mathrm{ex}e^{i(k_0+K_\mathrm{ex}/2)d}}}\abs{\frac{z_1^\mathrm{ex}}{z_2^\mathrm{ex}}}^m-1}} \abs{\frac{\partial_K z_2^\mathrm{ex}+iz_2^\mathrm{ex}d/2}{z_2^\mathrm{ex}}} \right)\equiv \mathcal{M}_{R,m}.
    \end{split}
\end{equation}
The majorizing sequence $\mathcal{M}_{R,m}$ has 
\begin{equation}
    \lim\limits_{m\to\infty} \mathcal{M}_{R,m} =\abs{\frac{\partial_K z_>^\mathrm{ex}+iz_>^\mathrm{ex}d/2}{z_>^\mathrm{ex}}},
\end{equation}
with
\begin{equation}
    \quad z_>^\mathrm{ex}=\begin{cases}
        z_1^\mathrm{ex}, & |z_1^\mathrm{ex}|\geq|z_2^\mathrm{ex}|,\\
        z_2^\mathrm{ex}, & |z_1^\mathrm{ex}|\leq|z_2^\mathrm{ex}|.
    \end{cases}
    \label{eq:zgex}
\end{equation}
Since $\mathcal{M}_{R,m}$ is convergent and $\mathcal{M}_{R,m}>0$, $\mathcal{M}_{R,m}$ is bounded from above. Hence, $|\tilde{\mu}_{R,m}|$ is also bounded. We denote $|\tilde{\mu}_{R,m}|\leq \mathcal{M}_R$, where $\mathcal{M}_R$ is the upper bound. $\hfill\blacksquare$

\subsubsection{2. Boundedness of $\tilde{\nu}_{R,m}$}
Likewise, we demonstrate that $|\tilde{\nu}_{R,m}|$ is bounded. The proof is also trivial for $z_1^\mathrm{ex}=z_2^\mathrm{ex}$, and we find
\begin{equation}
    \begin{split}
        |\tilde{\nu}_{R,m}|&\leq \frac{\varsigma S_\mathrm{ex}}{2}+\frac{1}{d} \left( \frac{1}{\abs{\abs{\frac{B_\mathrm{ex}}{A_\mathrm{ex}}\frac{1-z_1^\mathrm{ex}e^{i(k_0+K_\mathrm{ex}/2)d}}{1-z_2^\mathrm{ex}e^{i(k_0+K_\mathrm{ex}/2)d}}}\abs{\frac{z_2^\mathrm{ex}}{z_1^\mathrm{ex}}}^m-1}} \abs{\frac{e^{i\left(k_0+\frac{K_\mathrm{ex}}{2}\right)d}(\partial_K z_1^\mathrm{ex}+iz_1^\mathrm{ex}d/2)}{1-z_1^\mathrm{ex}e^{i(k_0+\frac{K_\mathrm{ex}}{2})d}}+\frac{\partial_K A_\mathrm{ex}}{A_\mathrm{ex}}} \right.\\
        &\left.\quad + \frac{1}{\abs{\abs{\frac{A_\mathrm{ex}}{B_\mathrm{ex}}\frac{1-z_2^\mathrm{ex}e^{i(k_0+K_\mathrm{ex}/2)d}}{1-z_1^\mathrm{ex}e^{i(k_0+K_\mathrm{ex}/2)d}}}\abs{\frac{z_1^\mathrm{ex}}{z_2^\mathrm{ex}}}^m-1}} \abs{\frac{e^{i\left(k_0+\frac{K_\mathrm{ex}}{2}\right)d}(\partial_K z_2^\mathrm{ex}+iz_2^\mathrm{ex}d/2)}{1-z_2^\mathrm{ex}e^{i(k_0+\frac{K_\mathrm{ex}}{2})d}}+\frac{\partial_K B_\mathrm{ex}}{B_\mathrm{ex}}} \right)\equiv \mathcal{N}_{R,m}.
    \end{split}
\end{equation}
for $z_1^\mathrm{ex}\neq z_2^\mathrm{ex}$. $\mathcal{N}_{R,m}$ has
\begin{equation}
    \lim\limits_{m\to\infty} \mathcal{N}_{R,m} = \frac{\varsigma S_\mathrm{ex}}{2}+\abs{\frac{e^{i\left(k_0+\frac{K_\mathrm{ex}}{2}\right)d}(\partial_K z_>^\mathrm{ex}+iz_>^\mathrm{ex}d/2)}{1-z_>^\mathrm{ex}e^{i(k_0+\frac{K_\mathrm{ex}}{2})d}}+\frac{\partial_K A_>^\mathrm{ex}}{A_>^\mathrm{ex}}},
\end{equation}
where 
\begin{equation}
    A_>^\mathrm{ex}=\begin{cases}
        A_\mathrm{ex}, & |z_1^\mathrm{ex}|\geq|z_2^\mathrm{ex}|,\\
        B_\mathrm{ex}, & |z_1^\mathrm{ex}|\leq|z_2^\mathrm{ex}|,
    \end{cases}
\end{equation}
and $z_>^\mathrm{ex}$ is the same as Eq.~\eqref{eq:zgex}. We thus know that $\mathcal{N}_{R,m}$ is bounded from above and $|\tilde{\nu}_{R,m}|$ is also bounded. We denote $|\tilde{\nu}_{R,m}|\leq \mathcal{N}_R$ with $\mathcal{N}_R$ the upper bound. $\hfill\blacksquare$

Similarly, the left-edge state can be expanded as
\begin{equation}
        \ket*{T_L}=i\frac{\Gamma_R}{2}\sigma_{k_0}^\dagger\sum\limits_{m=1}^{N-1} e^{i\delta Kd(m+m\tilde{\mu}_{L,m}+\tilde{\nu}_{L,m})} e^{i(K_\mathrm{ex}-k_0)md} L_m\sigma_m^\dagger \ket*{G} + \mathcal{O}(N^{-2}),
\end{equation}
with
\begin{subequations}
    \begin{align}
        \tilde{\mu}_{L,m}&=\frac{-i/d}{A_\mathrm{ex}\left[1-z_2^\mathrm{ex}e^{i(k_0-\frac{K_\mathrm{ex}}{2})d}\right]\left[z_1^\mathrm{ex}e^{i(k_0-\frac{K_\mathrm{ex}}{2})d}\right]^m+B_\mathrm{ex}\left[1-z_1^\mathrm{ex}e^{i(k_0-\frac{K_\mathrm{ex}}{2})d}\right]\left[z_2^\mathrm{ex}e^{i(k_0-\frac{K_\mathrm{ex}}{2})d}\right]^m} \notag \\
        &\quad \times\left( A_\mathrm{ex}\left[1-z_2^\mathrm{ex}e^{i(k_0-\frac{K_\mathrm{ex}}{2})d}\right]\left[z_1^\mathrm{ex}e^{i(k_0-\frac{K_\mathrm{ex}}{2})d}\right]^m \frac{\partial_K z_1^\mathrm{ex}-iz_1^\mathrm{ex}d/2}{z_1^\mathrm{ex}} \right. \notag\\
        &\left. \qquad +  B_\mathrm{ex}\left[1-z_1^\mathrm{ex}e^{i(k_0-\frac{K_\mathrm{ex}}{2})d}\right]\left[z_2^\mathrm{ex}e^{i(k_0-\frac{K_\mathrm{ex}}{2})d}\right]^m \frac{\partial_K z_2^\mathrm{ex}-iz_2^\mathrm{ex}d/2}{z_2^\mathrm{ex}} \right),\\
        \tilde{\nu}_{L,m}&=-\frac{i}{2}\varsigma S_\mathrm{ex} + \frac{-i/d}{A_\mathrm{ex}\left[1-z_2^\mathrm{ex}e^{i(k_0-\frac{K_\mathrm{ex}}{2})d}\right]\left[z_1^\mathrm{ex}e^{i(k_0-\frac{K_\mathrm{ex}}{2})d}\right]^m+B_\mathrm{ex}\left[1-z_1^\mathrm{ex}e^{i(k_0-\frac{K_\mathrm{ex}}{2})d}\right]\left[z_2^\mathrm{ex}e^{i(k_0-\frac{K_\mathrm{ex}}{2})d}\right]^m} \notag \\
        &\quad \times\left( A_\mathrm{ex}\left[1-z_2^\mathrm{ex}e^{i(k_0-\frac{K_\mathrm{ex}}{2})d}\right]\left[z_1^\mathrm{ex}e^{i(k_0-\frac{K_\mathrm{ex}}{2})d}\right]^m \left[\frac{e^{i\left(k_0-\frac{K_\mathrm{ex}}{2}\right)d}(\partial_K z_1^\mathrm{ex}-iz_1^\mathrm{ex}d/2)}{1-z_1^\mathrm{ex}e^{i(k_0-\frac{K_\mathrm{ex}}{2})d}}+\frac{\partial_K A_\mathrm{ex}}{A_\mathrm{ex}}\right] \right. \notag\\
        &\left. \qquad +  B_\mathrm{ex}\left[1-z_1^\mathrm{ex}e^{i(k_0-\frac{K_\mathrm{ex}}{2})d}\right]\left[z_2^\mathrm{ex}e^{i(k_0-\frac{K_\mathrm{ex}}{2})d}\right]^m \left[\frac{e^{i\left(k_0-\frac{K_\mathrm{ex}}{2}\right)d}(\partial_K z_2^\mathrm{ex}-iz_2^\mathrm{ex}d/2)}{1-z_2^\mathrm{ex}e^{i(k_0-\frac{K_\mathrm{ex}}{2})d}}+\frac{\partial_K B_\mathrm{ex}}{B_\mathrm{ex}}\right] \right),  
    \end{align}
\end{subequations}
and $L_m$ expressed in the main text. One can analogously prove that $\tilde{\mu}_{L,m}$ and $\tilde{\nu}_{L,m}$ are also bounded, so that the complex phase $\delta Kd(m+m\tilde{\mu}_{L,m}+\tilde{\nu}_{L,m})$ is of the order $\mathcal{O}(N^{-1})$ for a sufficiently large $N$. 

\subsection{D. Construction of $\mathbf{H}$\label{subsec:constructH}}
We present our motivation of the construction of the simpler Hamiltonian $\mathbf{H}$. To cancel the tails and determine the value of $K$, we expect the tails to be of the form $a_1(K)\ket*{\phi_1}+a_2(K)\ket*{\phi_2}$, where $a_{1,2}(K)$ are generic functions of $K$ and $\ket*{\phi_{1,2}}$ are global states independent of $K$. If so, these tails can be vanishing via the superposition of two equal-energy eigenstates. Nevertheless, $\ket*{T_{L,R}}$ do not exhibit the expected simple form.

To proceed, our treatment follows that the $K$-dependent phases of the sites that dominate the population of $\ket*{T_{L,R}}$ are slow-varying. For the right-edge state, it specifically means that $e^{i\delta Kd(N-m+1+m\tilde{\mu}_{R,m}+\tilde{\nu}_{R,m})}R_m$ varies slowly as $e^{i\delta KNd[1+\mathcal{O}(N^{-1})]}$ for $m\ll N$ as we have shown that $\delta Kd(-m+1+m\tilde{\mu}_{R,m}+\tilde{\nu}_{R,m})\sim\mathcal{O}(N^{-1})$, and has an exponentially small norm for $m\lesssim N$. Thus, we replace $e^{i\delta Kd(N-m+1+m\tilde{\mu}_{R,m}+\tilde{\nu}_{R,m})}$ by a \textit{mean-phase factor} $e^{i\delta KNd\mathbf{f}_R}$ with $\mathbf{f}_R=1+f_{R1}/N+\mathcal{O}(N^{-2})$ independent of $m$. Note that $\mathbf{f}_R$ is a function of $d$ and $\eta_{LR}$, and can be determined by the constraints that $\mathrm{Re}(\delta KNd\mathbf{f}_R)$ is the mean phase and $\mathrm{Im}(\delta KNd\mathbf{f}_R)$ ensures the same total population of $\ket*{T_R}$. Similarly, we can replace the factor $e^{i\delta Kd(m+m\tilde{\mu}_{L,m}+\tilde{\nu}_{L,m})}$ by the mean-phase factor $e^{i\delta KNd\mathbf{f}_L}$ with $\mathbf{f}_L=f_{L1}/N+\mathcal{O}(N^{-2})$. Note that in this work, we do not evaluate $\mathbf{f}_{L,R}$ directly for their analytical complicacy. With these simplifications, we can hence construct the simpler Hamiltonian $\mathbf{H}$ which has only the differences on the tails when acting on $\ket*{\psi_K}$, i.e., one defined by Eq.~\eqref{eq:simplerH} in the main text.

\subsection{E. Consistency of the perturbation theory\label{subsec:DeltaH}}
Here we provide the proof of $\mel*{\Psi}{\Delta H}{\Psi}\sim N^{-3}$. Following Eqs.~\eqref{eq:HeffonpsiK} and~\eqref{eq:simplerH} in the main text, we find that $\Delta H\ket*{\psi_{K_+}}=(\Delta H\ket*{\psi_{K_+}})_R+(\Delta H\ket*{\psi_{K_+}})_L$, where
\begin{equation}
    \begin{split}
        (\Delta H\ket*{\psi_{K_+}})_R&=-i\frac{\Gamma_L}{2} e^{i\pi\zeta\mathbf{f}_R}\sigma_{-k_0}^\dagger \sum\limits_{m=1}^{N-1} R_m e^{i(K_\mathrm{ex}+k_0)(N-m+1)d}\\
        &\quad \times\left[1-e^{i\delta K d(N-m+1+m\tilde{\mu}_{R,m}+\tilde{\nu}_{R,m}-N\mathbf{f}_R)}\right]\sigma_{N-m+1}^\dagger \ket*{G}+\mathcal{O}(N^{-2}),
    \end{split}
\end{equation}
and
\begin{equation}
    \begin{split}
        (\Delta H\ket*{\psi_{K_+}})_L&=-i\frac{\Gamma_R}{2} e^{i\pi\zeta\mathbf{f}_L}\sigma_{k_0}^\dagger \sum\limits_{m=1}^{N-1} L_m e^{i(K_\mathrm{ex}-k_0)md}\\
        &\quad \times\left[1-e^{i\delta K d(m+m\tilde{\mu}_{L,m}+\tilde{\nu}_{L,m}-N\mathbf{f}_L)}\right]\sigma_m^\dagger \ket*{G}+\mathcal{O}(N^{-2}).
    \end{split}
\end{equation}
Using Eq.~\eqref{eq:BSwf} in the main text, we obtain
\begin{equation}
    \begin{split}
        &\bra*{\Psi}(\Delta H\ket*{\psi_{K_+}})_R = -\frac{\Gamma_L}{\sqrt{2}N}e^{i\pi\zeta\mathbf{f}_R} \sum\limits_{\Delta/d=1}^{N-1}\sum\limits_{m=1}^{N-1} R_m C_\Delta^\mathrm{ex} (-1)^\zeta \left[1-e^{i\frac{\zeta\pi}{N}(m\tilde{\mu}_{R,m}+\tilde{\nu}_{R,m}+1-m-f_{R1})}\right]\\
        &\, \times \left(e^{-i\left(k_0+\frac{K_\mathrm{ex}}{2}\right)\Delta}\sin\left[\frac{\zeta\pi}{N}\left(1-m+\frac{\Delta}{2d}\right)\right] + e^{i\left(k_0+\frac{K_\mathrm{ex}}{2}\right)\Delta}\sin\left[\frac{\zeta\pi}{N}\left(1-m-\frac{\Delta}{2d}\right)\right]\right) + \text{(higher-order terms)}.
    \end{split}
\end{equation}
Here, we have omitted the higher-order terms. As $\tilde{\mu}_{R,m}$ and $\tilde{\nu}_{R,m}$ are bounded, the growth of the norm of $e^{i\frac{\zeta\pi}{N}(m\tilde{\mu}_{R,m}+\tilde{\nu}_{R,m}+1-m-f_{R1})}$ with $m$ is far overweighed by the decay of $R_m$. Thus, the terms to be summed are exponentially small for $m\lesssim N$. Meanwhile, all factors involving $\Delta$ are bounded apart from $C_\Delta^\mathrm{ex}$, so the summands are also exponentially suppressed for $\Delta/d\lesssim N$. Therefore, the leading-order contributions are the terms with $m\ll N$ and $\Delta/d\ll N$, where the Taylor expansions can be consistently performed. Consequently, we find that the summation scales as $N^{-2}$. Mathematically, We show the above argument as follows.

\subsubsection{The summation scales as $N^{-2}$}
Denote the target as $\mathcal{I}=\sum_{\Delta/d=1}^{N-1}\sum_{m=1}^{N-1} \mathscr{I}_{\Delta,m}$, with
\begin{equation}
    \begin{split}
        \mathscr{I}_{\Delta,m}=& R_m C_\Delta^\mathrm{ex} (-1)^\zeta \left[1-e^{i\frac{\zeta\pi}{N}(m\tilde{\mu}_{R,m}+\tilde{\nu}_{R,m}+1-m-f_{R1})}\right]\\
        &\, \times \left(e^{-i\left(k_0+\frac{K_\mathrm{ex}}{2}\right)\Delta}\sin\left[\frac{\zeta\pi}{N}\left(1-m+\frac{\Delta}{2d}\right)\right] + e^{i\left(k_0+\frac{K_\mathrm{ex}}{2}\right)\Delta}\sin\left[\frac{\zeta\pi}{N}\left(1-m-\frac{\Delta}{2d}\right)\right]\right).
    \end{split}
\end{equation}
We split the summations $\sum_{\Delta/d=1}^{N-1}=\sum_{\Delta/d=1}^{\lfloor \sqrt{N} \rfloor-1} + \sum_{\Delta/d=\lfloor \sqrt{N} \rfloor}^{N-1}$ and $\sum_{m=1}^{N-1}=\sum_{m=1}^{\lfloor \sqrt{N} \rfloor-1 } + \sum_{m=\lfloor \sqrt{N} \rfloor}^{N-1}$, and denote $\mathcal{I}=\mathcal{I}_1+\mathcal{I}_2+\mathcal{I}_3+\mathcal{I}_4$, with
\begin{subequations}
    \begin{align}
        \mathcal{I}_1 &= \sum_{\Delta/d=1}^{\lfloor \sqrt{N} \rfloor-1} \sum_{m=1}^{\lfloor \sqrt{N} \rfloor-1}\mathscr{I}_{\Delta,m},\quad
        \mathcal{I}_2 = \sum_{\Delta/d=\lfloor \sqrt{N} \rfloor}^{N-1} \sum_{m=1}^{\lfloor \sqrt{N} \rfloor-1}\mathscr{I}_{\Delta,m},\\
        \mathcal{I}_3 &= \sum_{\Delta/d=1}^{\lfloor \sqrt{N} \rfloor-1} \sum_{m=\lfloor \sqrt{N} \rfloor}^{N-1}\mathscr{I}_{\Delta,m},\quad
        \mathcal{I}_4 = \sum_{\Delta/d=\lfloor \sqrt{N} \rfloor}^{N-1} \sum_{m=\lfloor \sqrt{N} \rfloor}^{N-1}\mathscr{I}_{\Delta,m}.
    \end{align}
\end{subequations} 
Using the notations
\begin{subequations}
    \begin{align}
        \bar{R}_m&=|A_\mathrm{ex}|\frac{|z_1^\mathrm{ex}|^m}{1-|z_1^\mathrm{ex}|}+|B_\mathrm{ex}|\frac{|z_2^\mathrm{ex}|^m}{1-|z_2^\mathrm{ex}|},\\
        \bar{C}_\Delta^\mathrm{ex}&=|A_\mathrm{ex}||z_1^\mathrm{ex}|^{\Delta/d} + |B_\mathrm{ex}||z_2^\mathrm{ex}|^{\Delta/d},
    \end{align}
\end{subequations}
we have
\begin{equation}
    \begin{split}
        |\mathcal{I}_4| &\leq \sum_{\Delta/d=\lfloor \sqrt{N} \rfloor}^{N-1} \sum_{m=\lfloor \sqrt{N} \rfloor}^{N-1} 2\bar{R}_m \bar{C}_\Delta^\mathrm{ex} \left[ 1+e^{\frac{\zeta\pi}{N}(m\mathcal{M}_R+\mathcal{N}_R+|f_{R1}|)} \right]\\
        &\leq \sum_{\Delta/d=\lfloor \sqrt{N} \rfloor}^{\infty} \sum_{m=\lfloor \sqrt{N} \rfloor}^{\infty} 2\bar{R}_m \bar{C}_\Delta^\mathrm{ex} \left[ 1+e^{\zeta\pi\left(\mathcal{M}_R+\frac{\mathcal{N}_R+|f_{R1}|}{N}\right)} \right]\\
        &=2\left[ 1+e^{\zeta\pi\left(\mathcal{M}_R+\frac{\mathcal{N}_R+|f_{R1}|}{N}\right)} \right] \left( |A_\mathrm{ex}|\frac{|z_1^\mathrm{ex}|^{\lfloor \sqrt{N} \rfloor}}{1-|z_1^\mathrm{ex}|}+|B_\mathrm{ex}|\frac{|z_2^\mathrm{ex}|^{\lfloor \sqrt{N} \rfloor}}{1-|z_2^\mathrm{ex}|} \right) \left( |A_\mathrm{ex}|\frac{|z_1^\mathrm{ex}|^{\lfloor \sqrt{N} \rfloor}}{(1-|z_1^\mathrm{ex}|)^2}+|B_\mathrm{ex}|\frac{|z_2^\mathrm{ex}|^{\lfloor \sqrt{N} \rfloor}}{(1-|z_2^\mathrm{ex}|)^2} \right)\\
        &\sim \mathcal{O}\left(|z_>^\mathrm{ex}|^{2\sqrt{N}}\right),
    \end{split}
\end{equation}
which is exponentially small. Similarly, we find
\begin{subequations}
    \begin{align}
        |\mathcal{I}_3| & \leq2\left[ 1+e^{\zeta\pi\left(\mathcal{M}_R+\frac{\mathcal{N}_R+|f_{R1}|}{N}\right)} \right] \left( |A_\mathrm{ex}|\frac{|z_1^\mathrm{ex}|}{1-|z_1^\mathrm{ex}|}+|B_\mathrm{ex}|\frac{|z_2^\mathrm{ex}|}{1-|z_2^\mathrm{ex}|} \right) \left( |A_\mathrm{ex}|\frac{|z_1^\mathrm{ex}|^{\lfloor \sqrt{N} \rfloor}}{(1-|z_1^\mathrm{ex}|)^2}+|B_\mathrm{ex}|\frac{|z_2^\mathrm{ex}|^{\lfloor \sqrt{N} \rfloor}}{(1-|z_2^\mathrm{ex}|)^2} \right) \notag \\
        &\sim \mathcal{O}\left(|z_>^\mathrm{ex}|^{\sqrt{N}}\right),\\
        |\mathcal{I}_2| & \leq2\left[ 1+e^{\zeta\pi\left(\mathcal{M}_R+\frac{\mathcal{N}_R+|f_{R1}|}{N}\right)} \right] \left( |A_\mathrm{ex}|\frac{|z_1^\mathrm{ex}|^{\lfloor \sqrt{N} \rfloor}}{1-|z_1^\mathrm{ex}|}+|B_\mathrm{ex}|\frac{|z_2^\mathrm{ex}|^{\lfloor \sqrt{N} \rfloor}}{1-|z_2^\mathrm{ex}|} \right) \left( |A_\mathrm{ex}|\frac{|z_1^\mathrm{ex}|}{(1-|z_1^\mathrm{ex}|)^2}+|B_\mathrm{ex}|\frac{|z_2^\mathrm{ex}|}{(1-|z_2^\mathrm{ex}|)^2} \right) \notag \\
        &\sim \mathcal{O}\left(|z_>^\mathrm{ex}|^{\sqrt{N}}\right).
    \end{align}
\end{subequations}
For $\mathcal{I}_1$, on the other hand, since $\frac{\Delta/d}{N}, \frac{m}{N}\leq \frac{1}{\sqrt{N}}$, $\frac{|m\tilde{\mu}_{R,m}|}{N}\leq \frac{\mathcal{M}_R}{\sqrt{N}}$ and $\frac{|\tilde{\nu}_{R,m}|}{N}\leq \frac{\mathcal{N}_R}{\sqrt{N}}$, 
\begin{equation}
    \frac{\Delta/d}{N},\frac{m}{N},\frac{m\tilde{\mu}_{R,m}}{N}\text{ and }\frac{\tilde{\nu}_{R,m}}{N}\to 0
\end{equation}
holds uniformly as $N\to\infty$. This allows us to expand $\mathcal{I}_1$ as
\begin{equation}
    \begin{split}
        \mathcal{I}_1&=\sum_{\Delta/d=1}^{\lfloor \sqrt{N} \rfloor-1} \sum_{m=1}^{\lfloor \sqrt{N} \rfloor-1} R_m C_\Delta^\mathrm{ex}(-1)^\zeta \left( e^{-i\left(k_0+\frac{K_\mathrm{ex}}{2}\right)\Delta}\frac{\zeta\pi}{N}\left(1-m+\frac{\Delta}{2d}\right)+\mathcal{O}\left[\left(\frac{m-\Delta/d}{N}\right)^2\right]   \right.\\
        &\qquad\qquad\qquad\qquad\qquad\qquad \left. +  e^{i\left(k_0+\frac{K_\mathrm{ex}}{2}\right)\Delta}\frac{\zeta\pi}{N}\left(1-m-\frac{\Delta}{2d}\right)+\mathcal{O}\left[\left(\frac{m+\Delta/d}{N}\right)^2\right]  \right)\\
        &\quad\times\left(-i\frac{\zeta\pi}{N}(m\tilde{\mu}_{R,m}+\tilde{\nu}_{R,m}+1-m-f_{R1})+\mathcal{O}\left[\left(\frac{m\tilde{\mu}_{R,m}+\tilde{\nu}_{R,m}-m}{N}\right)^2\right]\right)
    \end{split}
\end{equation}
and to obtain 
\begin{equation}
    \begin{split}
        |\mathcal{I}_1|&\leq \frac{\zeta^2\pi^2}{N^2}\sum_{\Delta/d=1}^{\infty} \sum_{m=1}^{\infty} 2 \bar{R}_m \bar{C}_\Delta^\mathrm{ex}\left(1+m+\frac{\Delta}{2d}\right)[(\mathcal{M}_R+1)m+\mathcal{N}_R+1+|f_{R1}|]+\mathcal{O}\left(\frac{1}{N^3}\right)\\
        &\sim \mathcal{O}\left(\frac{1}{N^2}\right).
    \end{split}
\end{equation}
Hence, we find $|\mathcal{I}|\leq\mathcal{O}(N^{-2})+(\text{exponentially small terms})$, i.e., the summation $\mathcal{I}\sim N^{-2}$. $\hfill\blacksquare$

We adopt the short-hand notation $\sum_{\Delta/d\ll N}\sum_{m\ll N}$ to represent the dominant contribution ($\mathcal{I}_1$) of $\mathcal{I}$, so that
\begin{equation}
    \begin{split}
        \bra*{\Psi}(\Delta H\ket*{\psi_{K_+}})_R &=i\frac{\pi^2\zeta^2\Gamma_L}{\sqrt{2}N^3} \sum\limits_{\Delta/d\ll N}\sum\limits_{m\ll N} R_m C_\Delta^\mathrm{ex}  (m\tilde{\mu}_{R,m}+\tilde{\nu}_{R,m}+1-m-f_{R1})\\
        & \quad \times \left[e^{-i\left(k_0+\frac{K_\mathrm{ex}}{2}\right)\Delta}\left(1-m+\frac{\Delta}{2d}\right) + e^{i\left(k_0+\frac{K_\mathrm{ex}}{2}\right)\Delta}\left(1-m-\frac{\Delta}{2d}\right)\right] + \mathcal{O}(N^{-4}).
    \end{split}
\end{equation} 
Similarly, the contribution from the left-edge part is
\begin{equation}
    \begin{split}
        \bra*{\Psi}(\Delta H\ket*{\psi_{K_+}})_L &= -\frac{\Gamma_R}{\sqrt{2}N}e^{i\pi\zeta\mathbf{f}_L} \sum\limits_{\Delta/d=1}^{N-1}\sum\limits_{m=1}^{N-1} L_m C_\Delta^\mathrm{ex} \left[1-e^{i\frac{\zeta\pi}{N}(m\tilde{\mu}_{L,m}+\tilde{\nu}_{L,m}+m-f_{L1})}\right]\\
        &\times \left(e^{i\left(k_0-\frac{K_\mathrm{ex}}{2}\right)\Delta}\sin\left[\frac{\zeta\pi}{N}\left(m+\frac{\Delta}{2d}\right)\right] + e^{-i\left(k_0-\frac{K_\mathrm{ex}}{2}\right)\Delta}\sin\left[\frac{\zeta\pi}{N}\left(m-\frac{\Delta}{2d}\right)\right]\right) + \text{(higher-order terms)}\\
        &=i\frac{\pi^2\zeta^2\Gamma_R}{\sqrt{2}N^3} \sum\limits_{\Delta/d\ll N}\sum\limits_{m\ll N} L_m C_\Delta^\mathrm{ex} (m\tilde{\mu}_{L,m}+\tilde{\nu}_{L,m}+m-f_{L1})\\
        & \quad \times \left[e^{i\left(k_0-\frac{K_\mathrm{ex}}{2}\right)\Delta}\left(m+\frac{\Delta}{2d}\right) + e^{-i\left(k_0-\frac{K_\mathrm{ex}}{2}\right)\Delta}\left(m-\frac{\Delta}{2d}\right)\right] + \mathcal{O}(N^{-4}).
    \end{split}
\end{equation} 
Following the derivation above, we find that $\bra*{\Psi}\Delta H\ket*{\psi_{K_-}}=\bra*{\Psi}(\Delta H\ket*{\psi_{K_-}})_R+\bra*{\Psi}(\Delta H\ket*{\psi_{K_-}})_L$, with $\bra*{\Psi}(\Delta H\ket*{\psi_{K_-}})_{L,R}=-\bra*{\Psi}(\Delta H\ket*{\psi_{K_+}})_{L,R}$ to the leading order. Finally, we obtain
\begin{equation}
    \begin{split}
        \mel*{\Psi}{\Delta H}{\Psi}&=\frac{1}{\sqrt{2}}\left[\bra*{\Psi}(\Delta H\ket*{\psi_{K_+}})_L-\bra*{\Psi}(\Delta H\ket*{\psi_{K_-}})_L\right] + \frac{1}{\sqrt{2}}\left[\bra*{\Psi}(\Delta H\ket*{\psi_{K_+}})_R-\bra*{\Psi}(\Delta H\ket*{\psi_{K_-}})_R\right]\\
        &\approx \sqrt{2}\left[\bra*{\Psi}(\Delta H\ket*{\psi_{K_+}})_L+\bra*{\Psi}(\Delta H\ket*{\psi_{K_+}})_R\right] \sim N^{-3},
    \end{split}
\end{equation}
which is an order smaller than the separation of the eigenenergy $\mathcal{E}_\zeta\approx\mathcal{E}_\mathrm{ex}+\pi^2\zeta^2\alpha_2/N^2d^2$. Therefore, $\Delta H$ is consistently a perturbation to $\mathbf{H}$.
\end{widetext}

\section{S-IV. Characterization of two-excitation eigenstates\label{sec:Character}}
A detailed numerical characterization of the two-excitation eigenstates is presented in this section, where the BSs are especially highlighted. 

\subsection{A. Numerical identification of states}
\begin{figure}[htbp]
    \includegraphics[scale=0.56]{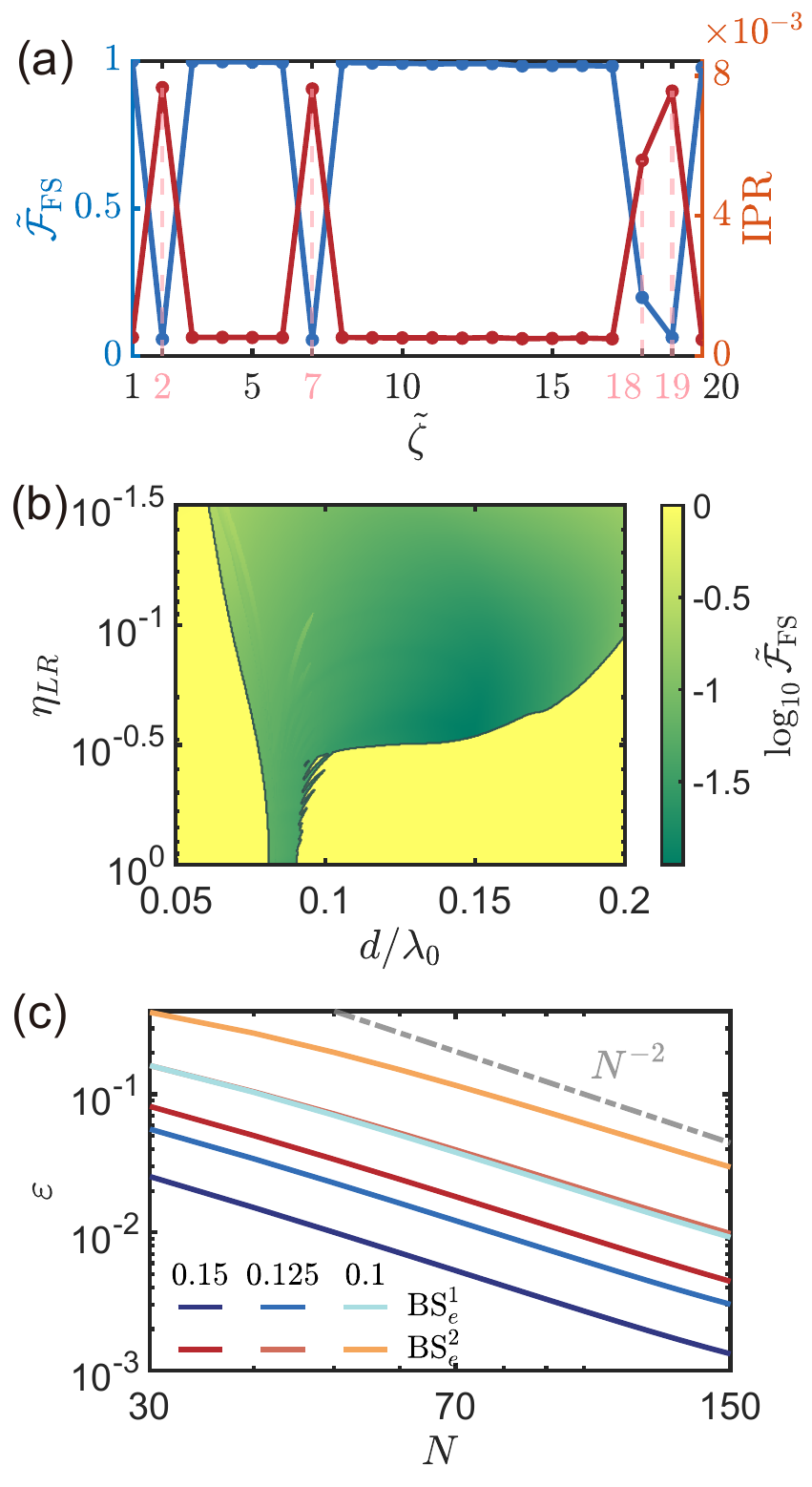}
    \centering
    \caption{(a) Fermionic overlap (blue) and IPR (red) for the $20$ most subradiant states, with $N=100$, $d=0.15\lambda_0$ and $\eta_{LR}=10^{-0.5}$ as an example. (b) Fermionic overlap distribution with respect to $d$ and $\eta_{LR}$ for $N=100$. The solid curve shows the boundary of the FS region. (c) Scaling of the infidelity between the numerical BS and the ansatz for fixed $\eta_{LR}=10^{-0.75}$, where the states and the values of $d/\lambda_0$ are specified by the legend. The dashed dotted line represents the reference $N^{-2}$ dependence.}
    \label{fgS:identify} 
\end{figure}

\begin{figure}[htbp]
    \includegraphics[scale=0.52]{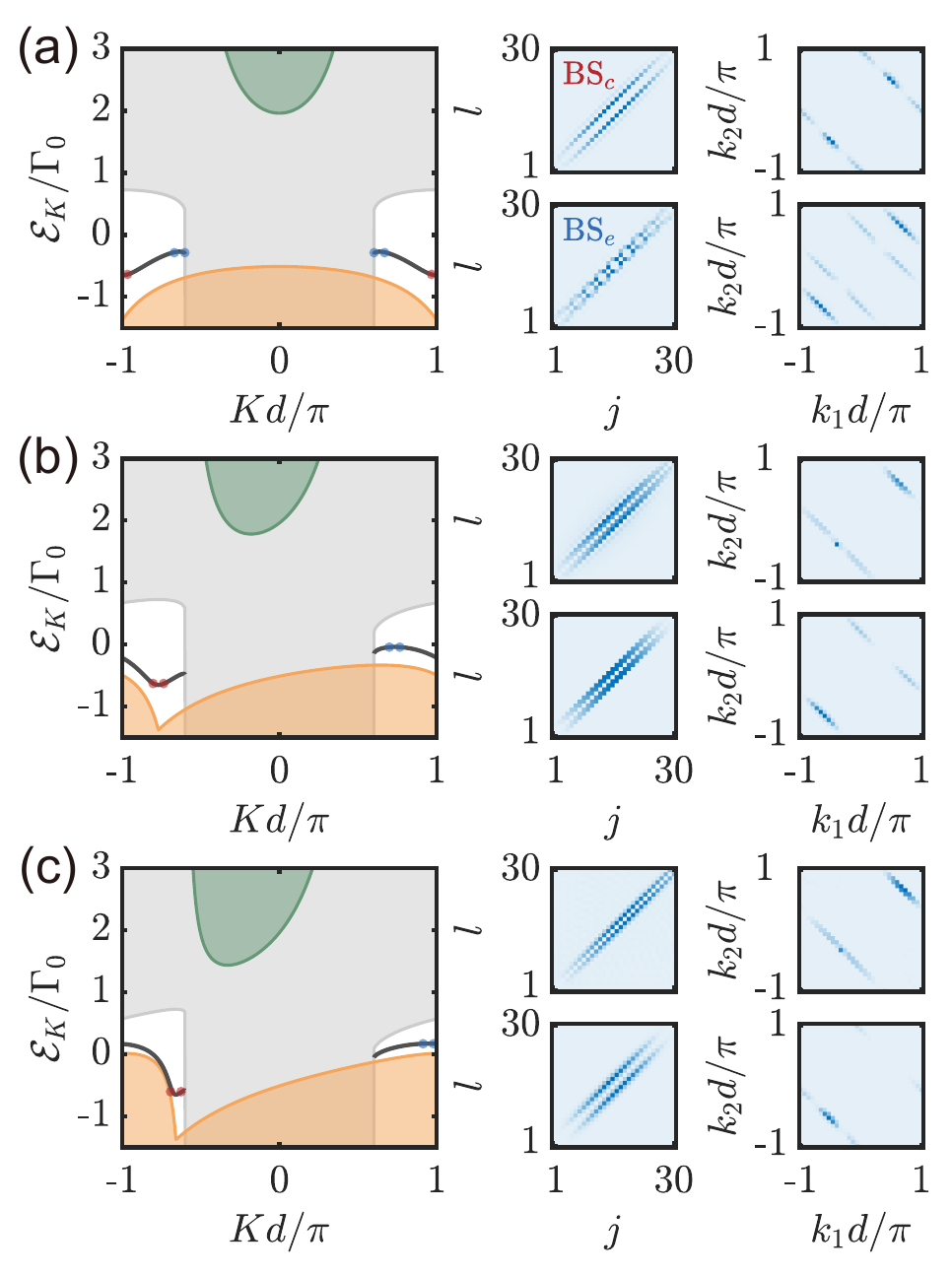}
    \centering
    \caption{Left panels: two-excitation dispersion relation for an infinite array as a function of the center-of-excitation momentum. The green (orange) shaded area represents the continuum originating from two upper-branch (lower-branch) polaritons, while the grey shaded area is accessed by the combination of the polaritons from the upper and the lower branches. The black solid line in the gap shows the BS energy band, where the circles correspond to $\mathrm{BS}_c$ (red) and $\mathrm{BS}_e$ (blue) for an array of $N=30$. Right panels: real (left) and momentum (right) space probability distributions of $\mathrm{BS}_c$ (upper) and $\mathrm{BS}_e$ (lower) for $N=30$. Here, all the plots have $d=0.15\lambda_0$, with $\eta_{LR}=1$ (a), $10^{-0.5}$ (b) and $10^{-1}$ (c), respectively.} 
    \label{fgS:spectrum} 
\end{figure}

We identify FSs and BSs among the numerically obtained eigenstates as follows. First, since the FSs has a relatively simpler probability distribution independent of $\eta_{LR}$ (for a sufficiently large $N$), we generate the (normalized) FS ansatz
\begin{equation}
    |c_{\xi_1\xi_2,jl}^{\text{(FS)}}|\propto \sin\frac{\xi_1\pi j}{N}\sin\frac{\xi_2\pi l}{N}-\sin\frac{\xi_2\pi j}{N}\sin\frac{\xi_1\pi l}{N}.
\end{equation}
For the $\tilde{\zeta}$th most subradiant eigenstate $\ket*{\psi_{\tilde{\zeta}}}=\sum_{j<l} c_{{\tilde{\zeta}},jl}\sigma_j^\dagger\sigma_l^\dagger\ket*{G}$, we determine whether it is an FS by evaluating the Fermionic overlap
\begin{equation}
    \tilde{\mathcal{F}}_{\mathrm{FS}}=\max\limits_{\xi_1,\xi_2} \sum\limits_{j<l}^N |c_{\xi_1\xi_2,jl}^{\text{(FS)}}| |c_{\tilde{\zeta},jl}|.
\end{equation}
Note that the eigenstate is also normalized by $\sum_{j<l}|c_{\tilde{\zeta},jl}|^2=1$. As an example, we show the results for the $20$ most subradiant states in Fig.~\hyperref[fgS:identify]{S4(a)}, with $N=100$, $d=0.15\lambda_0$ and $\eta_{LR}=10^{-0.5}$. Apart from the dips occurred at $\tilde{\zeta}=2,7,18,19$, the values of $\tilde{\mathcal{F}}_{\mathrm{FS}}$ are greater than $0.975$, indicating that the states with $\tilde{\mathcal{F}}_{\mathrm{FS}}$ close to $1$ are indeed FSs. Meanwhile, we compute the inverse participation ratio (IPR)
\begin{equation}
    \mathrm{IPR}=\sum\limits_{j<l}^N|c_{\tilde{\zeta},jl}|^4
\end{equation}
for these eigenstates. The non-FS states here possess larger $\mathrm{IPR}$, suggesting that they are more spatially localized. As a natural second step, we check whether the non-FS states are BSs by computing the fidelity with the BS ansatz:
\begin{equation}
    \mathcal{F}_\mathrm{BS}=\max\limits_{K_\mathrm{ex},\zeta} |\braket*{\Psi_\zeta}{\psi_{\tilde{\zeta}}}|^2. 
\end{equation}
Here, $\ket*{\Psi_\zeta}$ is given by Eq.~\eqref{eq:BSwf} in the main text. We find $\mathcal{F}_\mathrm{BS}=0.9979, 0.9915, 0.9926, 0.9773$ for $\tilde{\zeta}=2,7,18,19$, respectively. Therefore, these states are exactly BSs. More specifically, they are $\mathrm{BS}_e^{\zeta=1},\mathrm{BS}_e^{\zeta=2},\mathrm{BS}_c^{\zeta=1}$ and $\mathrm{BS}_e^{\zeta=3}$ [cf. Figs.~\hyperref[fgS:spectrum]{S5(b)} and~\hyperref[fgS:BS1to6]{S7(b)}], respectively.

The results above imply that a subradiant eigenstate is a BS if it is not an FS for $\tilde{\zeta}\ll N$. Thus, in the extensive calculations performed over $d$ and $\eta_{LR}$, FS and BS are identified solely by computing $\tilde{\mathcal{F}}_{\mathrm{FS}}$. In the $d-\eta_{LR}$ plane for the decay rates of the most subradiant states, there exists a well-defined FS region where $\tilde{\mathcal{F}}_{\mathrm{FS}}$ exceeds $0.98$ throughout, which is displayed in Fig.~\hyperref[fgS:identify]{S4(b)}. We hence deduce that the boundary of this region is the FS-BS separation line, as plotted in Fig.~\hyperref[fg:scenario]{1(c)} of the main text.

We further check the BS ansatz by evaluating the infidelity between the numerical eigenstate and the ansatz wavefunction. The scaling of the infidelity $\varepsilon=1-\mathcal{F}_\mathrm{BS}$ is shown in Fig.~\hyperref[fgS:identify]{S4(c)}, for both $\mathrm{BS}_e^{1}$ and $\mathrm{BS}_e^{2}$ with $d/\lambda_0=0.1,0.125,0.15$ and fixed $\eta_{LR}=10^{-0.75}$. The infidelity scales as $N^{-2}$ for sufficiently large $N$, indicating that the $\mathcal{O}(N^{-1})$ residue of Eq.~\eqref{eq:BSwf} in the main text is orthorgonal to the dominant component.

\subsection{B. Two-excitation dispersion relation}
To further show the impact of chirality on the energy spectrum, we plot in Fig.~\ref{fgS:spectrum} the two-excitation dispersion along with the continuum for infinite systems. We choose $\eta_{LR}=1,10^{-0.5}$ and $10^{-1}$ for Figs.~\hyperref[fgS:spectrum]{S5(a)}, \hyperref[fgS:spectrum]{S5(b)} and~\hyperref[fgS:spectrum]{S5(c)}, while the spacing $d=0.15\lambda_0$ is fixed. In each case, the BS band lies within the gap of the continuum, where the continuum has
\begin{equation}
    \mathcal{E}_K^{(\text{con})}=\omega_{q}+\omega_{K-q},\quad -\frac{\pi}{d}\leq q\leq\frac{\pi}{d}.
\end{equation}
As the chirality increases, this gap progressively narrows, resulting in the reduction of $|\alpha_{2e}|$ of the BS dispersion.

The probability distributions of the corresponding BSs are also shown, in both the real space and the momentum space (see Fig.~\ref{fgS:spectrum}). Here, the Fourier transform of a state $\ket*{\psi}=\sum_{j<l}^N c_{jl}\sigma_j^\dagger\sigma_l^\dagger\ket*{G}$ is
\begin{equation}
    \tilde{c}_{k_1 k_2}=\sum\limits_{j<l}^N c_{jl} e^{-ik_1x_j-ik_2x_l}.
\end{equation}
Most BSs feature two distinct sets of well-defined iso-momentum lines in the $k_1-k_2$ plane. However, the $\mathrm{BS}_e$-like state exhibits four iso-momentum lines only for $\eta_{LR}=1$, which is a hybrid of the states near two degenerate extrema. We omit this exception in the decay rate computation in Fig.~\ref{fg:BSeAlpha2} of the main text.

\subsection{C. Subradiant decay rates}
\begin{figure}[htbp]
    \includegraphics[scale=0.57]{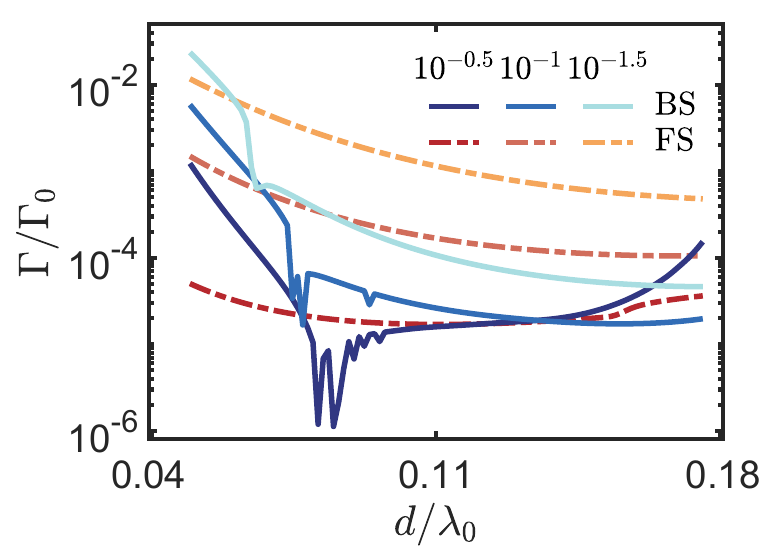}
    \centering
    \caption{Decay rates of the most subradiant BSs (solid curves) and FSs (dashed dotted curves) as a function of the atomic spacing. The results are obtained for an $N=100$ array, with $\eta_{LR}=10^{-0.5}$, $10^{-1}$ and $10^{-1.5}$, respectively.} 
    \label{fgS:scand} 
\end{figure}

Besides the subradiant decay rates for varying $\eta_{LR}$ in Fig.~\hyperref[fg:BSeAlpha2]{2(a)}, figure.~\ref{fgS:scand} shows the dependence of the decay rates on $d$, for both FSs and BSs with $N=100$. Note that we do not distinguish the specific type of the BS, but rather focus on the most subradiant one for each given parameter set. The results for the FSs are the same as those in Fig.~\hyperref[fgS:FSdeta]{S2(b)}.

\begin{figure}[htbp]
    \includegraphics[scale=0.57]{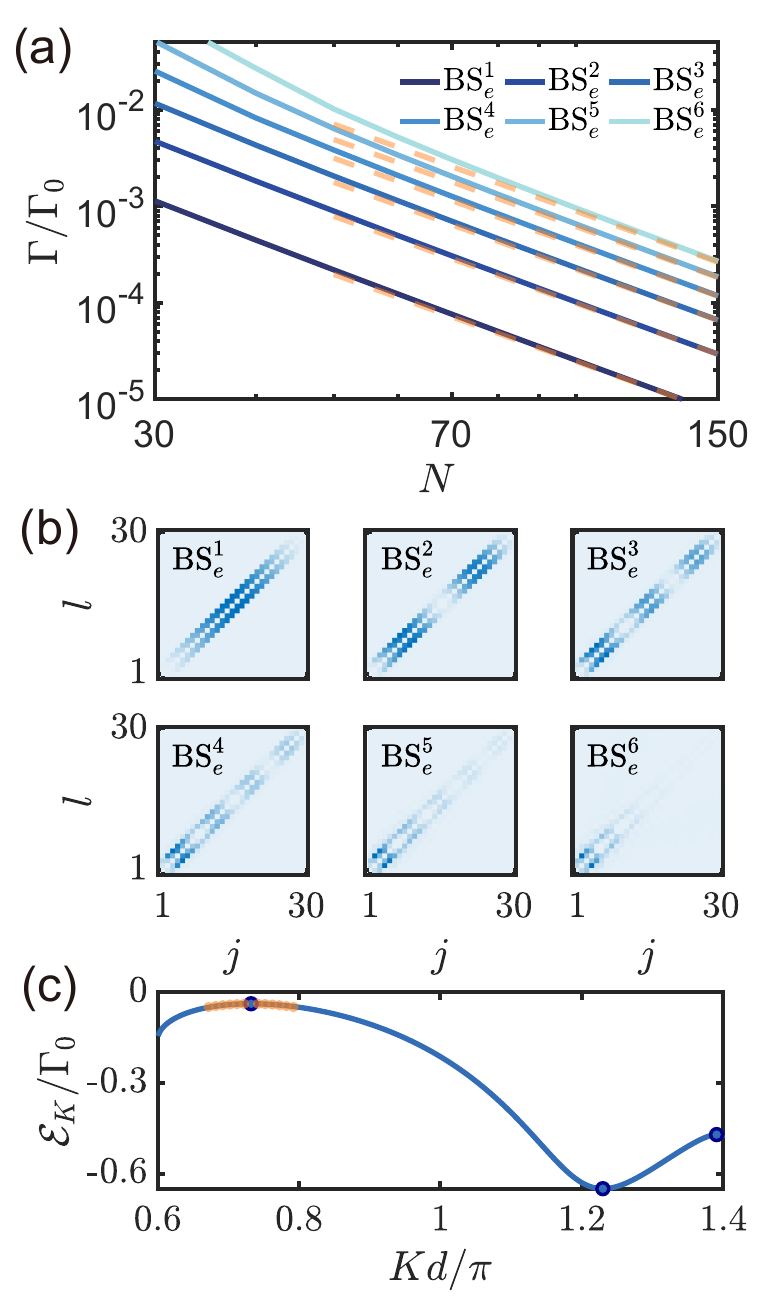}
    \centering
    \caption{(a) Decay rate scaling of $\mathrm{BS}_e^\zeta$ with $\zeta=1,2,\dots,6$, as indicated by the legend. The dashed lines are the semi-analytical results plotted by $\Gamma_\zeta=\zeta^2 |\alpha_{2e}|\mathbf{s}_e/N^3$. (b) Probability distributions of $\mathrm{BS}_e^1,\dots,\mathrm{BS}_e^6$ for $N=30$, with the corresponding eigenenergies (orange circles) shown in (c). Computations here are performed for $d=0.15\lambda_0$ and $\eta_{LR}=10^{-0.5}$.} 
    \label{fgS:BS1to6} 
\end{figure}

Moreover, we also demonstrate the $\zeta^2$ dependence of the BS decay rates predicted by Eq.~\eqref{eq:BSscaling} in the main text. In Fig.~\hyperref[fgS:BS1to6]{S7(a)}, we find that the $\mathrm{BS}_e^\zeta$ decay rates scale as $\Gamma_\zeta\sim\zeta^2/N^3$ for $d=0.15\lambda_0$ and $\eta_{LR}=10^{-0.5}$, with the numbers of anitnodes $\zeta=1,2,\dots,6$. This dependence is highlighted by the semi-analytical curves using Eq.~\eqref{eq:BSscaling} of the main text, where the value of $\mathbf{s}_e$ is deduced from $\mathbf{s}_e=\Gamma_{\zeta=1} N^3/|\alpha_{2e}|$ for $\mathrm{BS}_e^1$ at $N=150$. Meanwhile, the states $\mathrm{BS}_e^\zeta$ are shown explicitly in Fig.~\hyperref[fgS:BS1to6]{S7(b)}. The left-concentrated feature wherein similar to Ref.~\cite{shi2025chiral} is a finite-size effect. The eigenenergies are also displayed on the BS band in Fig.~\hyperref[fgS:BS1to6]{S7(c)}.

\subsection{D. Stronger chirality}
\begin{figure}[htbp]
    \includegraphics[scale=0.57]{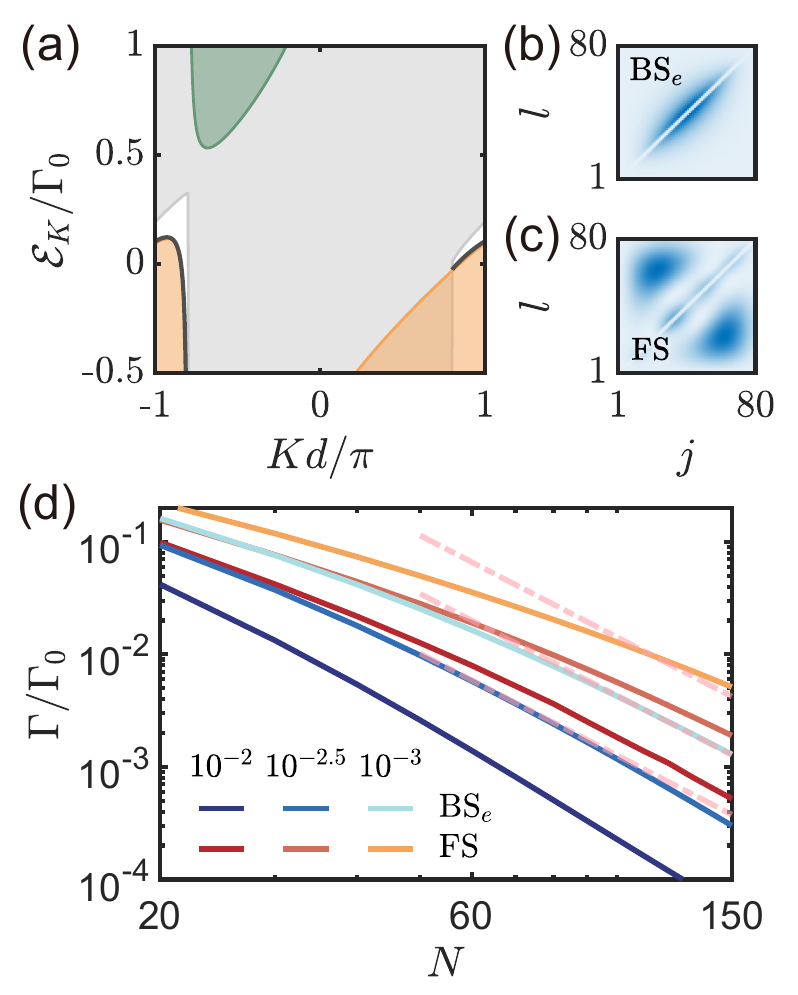}
    \centering
    \caption{(a) Two-excitation dispersion relation for $d=0.2\lambda_0$ and $\eta_{LR}=10^{-2}$. The notations of areas and curves are the same as Fig.~\hyperref[fgS:spectrum]{S5(a)}. Here we show the probability distributions of $\mathrm{BS}_e$ and FS in panels (b) and (c), respectively, for an array of $N=80$ and other parameters same as (a). (d) Solid curves demonstrate the numrical decay rate scaling, where the states and the values of $\eta_{LR}$ are specified by the legend. Dashed dotted lines represent the analytical results for FSs.} 
    \label{fgS:strong} 
\end{figure}

Computations for two-excitation subradiant states with $N=100$ in the main text are restricted to $\eta_{LR}\geq 10^{-1.5}$. When the chirality is even stronger, e.g., $\eta_{LR}=10^{-2}$, we find in Fig.~\hyperref[fgS:strong]{S8(a)} that the BS dispersion lies extremely close to the continuum. Therefore, the BSs and FSs for a finite array hybridize with each other, which can be found in Figs.~\hyperref[fgS:strong]{S8(b)} and~\hyperref[fgS:strong]{S8(c)}. Figure~\hyperref[fgS:strong]{S8(d)} shows that decay rates of these states also scale as $N^{-3}$ for large $N$. Nevertheless, the state hybridization leads to a dramatic discrepancy in FS decay rates between the numerical results and the analytical predictions.

\subsection{E. Robustness against disorder}
\begin{figure}[htbp]
    \includegraphics[scale=0.57]{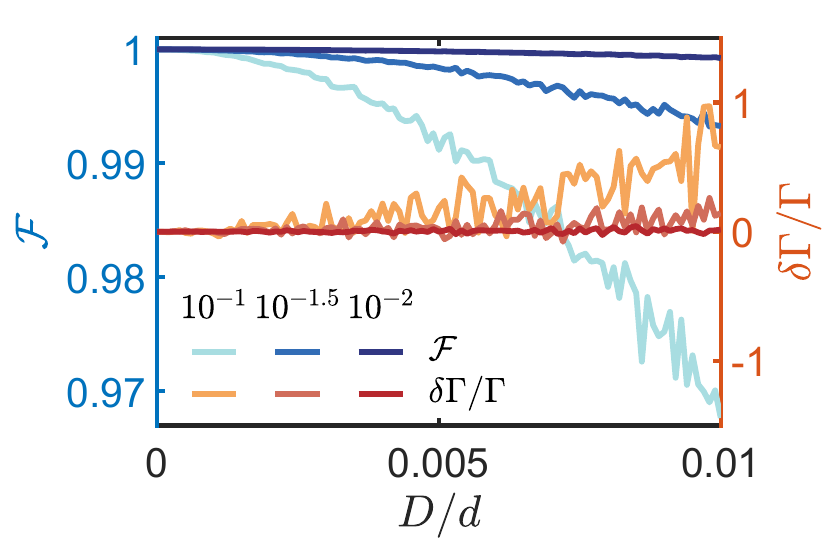}
    \centering
    \caption{Fidelity (blue) and decay rate difference (red) as a function of $D$. Parameters used for the calculations are $N=50$ and $d=0.15\lambda_0$, with the chirality $\eta_{LR}=10^{-1}, 10^{-1.5}$ and $10^{-2}$, respectively. The results are averaged over $N_s=200$ samples of disordered arrays.} 
    \label{fgS:disorder} 
\end{figure}

Here we study the robustness of BSs against weak spatial disorder. We introduce a Gaussian random coordinate fluctuation $\delta x_j$ which obeys the probability distribution
\begin{equation}
    f(\delta x_j)=\frac{1}{\sqrt{2\pi}D} e^{-\frac{(\delta x_j)^2}{2D^2}},
\end{equation}
so that the position of the $j$th atom is shifted to $x_j^{(\text{dis})}=jd+\delta x_j$. To characterize the impact of disorder, we define the fidelity 
\begin{equation}
    \mathcal{F}=|\braket*{\psi_{{\tilde{\zeta}}=1}^{(\text{dis})}}{\psi_{{\tilde{\zeta}}=1}}|^2,
\end{equation}
where $\ket*{\psi_{{\tilde{\zeta}}=1}}$ is the most subradiant state without disorder (which is also a BS here), and $\ket*{\psi_{{\tilde{\zeta}}=1}^{(\text{dis})}}$ is the most subradiant one for a disordered array. The fidelity decay with the increase of $D$ is shown in Fig.~\ref{fgS:disorder}. Notably, stronger chirality leads to gentler fidelity decay. We also show the difference of the minimal decay rates between the disordered and the disorder-free models. The results indicate that arrays with
stronger chirality are more insensitive to spatial disorder.

\subsection{F. Interaction-induced localized states}
\begin{figure}[htbp]
    \includegraphics[scale=0.58]{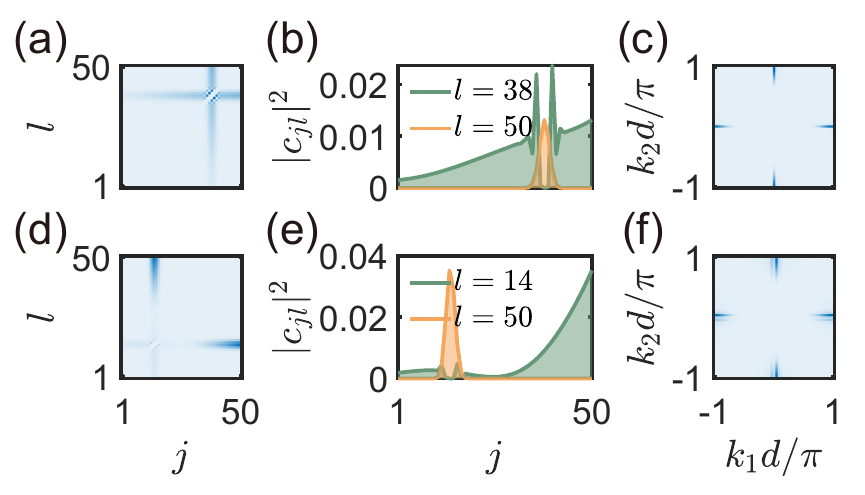}
    \centering
    \caption{Real space probability distributions of the IIL states with $\tilde{\zeta}=1145$ (a) and $1203$ (d). The distributions for fixed index $l$ (b,e) and in the momentum space (c,f) of these two states are also presented. The computation parameters here are $N=50$, $d=0.01\lambda_0$ and $\eta_{LR}=10^{-0.5}$.} 
    \label{fgS:IIL} 
\end{figure}

Apart from the FSs and BSs in the subradiant regime, we also observe another distinct class of eigenstates. In the strongly subwavelength limit with $d\ll \lambda_0$, we find in Fig.~\ref{fgS:IIL} the superradiant states that resemble the interaction-induced localized (IIL) states in Ref.~\cite{zhong2020photon,sheremet2023waveguide}. For a chiral array, the spatial distributions of these states do not cross at the center, but exhibit offsets to either the left or the right. Studying the origin and the radiative properties of the chiral IIL states may be an intriguing future direction.

\section{S-V. A simpler toy model\label{sec:toy}}
\begin{figure}[htbp]
    \includegraphics[scale=0.515]{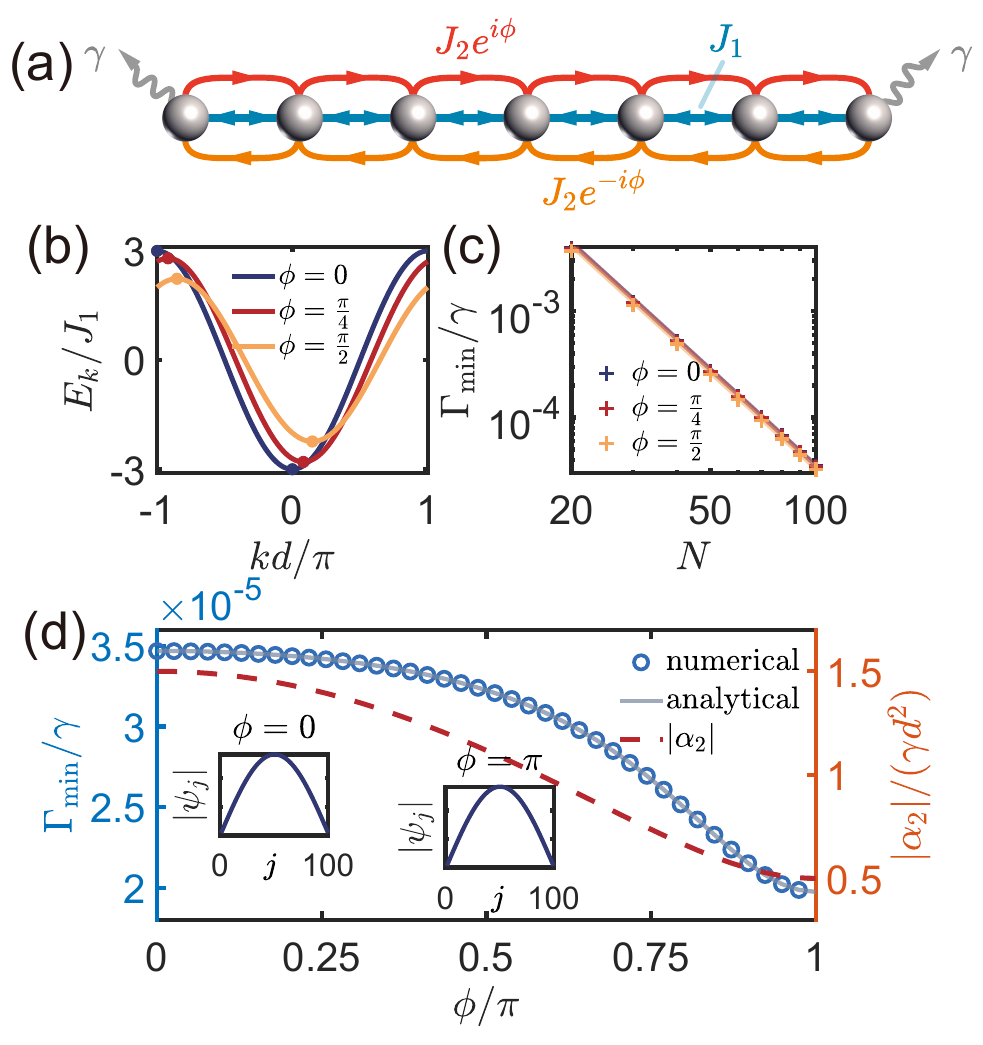}
    \centering
    \caption{(a) Schematic of the toy model. (b) Single-excitation dispersion relations for infinite arrays with $\phi=0, \pi/4$ and $\pi/2$. Circles mark the extremum points of each curve. (c) Decay rate scaling of the most subradiant states. Crosses (solid lines) represent the numerical (analytical) results. (d) Minimal decay rates (circles: numerical, solid curve: analytical) and $|\alpha_2|$ (dashed curve) as a function of $\phi$, with fixed $N=100$. The insets shows the probability amplitude distributions of the most subradiant eigenstates for $\phi=0$ (left) and $\phi=\pi$ (right). For all computations here, parameters $J_1=2J_2=\gamma$ are used.} 
    \label{fgS:toy} 
\end{figure}

In this section, we present a supplemental model to display that the subradiance mechanism regarding chirality would be general. Motivated by Ref.~\cite{shi2025chiral}, we consider a one-dimensional tight-binding model:
\begin{equation}
    \begin{split}
        H_a&=-\sum\limits_{j=1}^{N-1}(J_1+J_2e^{i\phi})a_{j+1}^\dagger a_j + \text{h.c.}\\
        &\quad -i\frac{\gamma}{2}(a_1^\dagger a_1+a_N^\dagger a_N),
    \end{split}
\end{equation}
which is illustrated in Fig.~\hyperref[fgS:toy]{S11(a)}. Here, $a_j$ ($a_j^\dagger$) is the Bosonic annihilation (creation) operator on the $j$th site. The hopping consists a non-chiral part with strength $J_1$ and a chiral part characterized by strength $J_2$ and phase $\phi$. Both $J_1$ and $J_2$ are real numbers. We introduce the parameters $t=(J_1^2+J_2^2+2J_1 J_2\cos\phi)^{1/2}$ and $\theta=\tan^{-1}[J_2\sin\phi/(J_1+J_2\cos\phi)]$ that satisfy $J_1+J_2e^{i\phi}=te^{i\theta}$ for simplicity. To achieve non-vanishing radiative decay of the eigenstates, the on-site dissipation with decay rate $\gamma$ on the $1$st and $N$th sites are also included.

In the single-excitation sector of $H_a$, the eigenstates are Bloch states $\ket*{k}$ with the dispersion $E_k=-2t\cos(kd-\theta)$ (where $d$ is the lattice constant) for an infinite chain. We plot in Fig.~\hyperref[fgS:toy]{S11(b)} the dispersion relations for $\phi=0,\pi/4$ and $\pi/2$, where the coupling constants have $J_1=2J_2=\gamma$. Tuning up the chirality parameter $\phi$, the value of $|\alpha_2|$ for the extrema can experience a decrease in this model, as shown more explicitly in Fig.~\hyperref[fgS:toy]{S11(d)}. 

For a finite chain, we denote the eigenstate as $\ket*{\psi}=\sum_{j=1}^N \psi_j a_j^\dagger\ket*{G}$, which satisfies the eigenequation $H_a\ket*{\psi}=E\ket*{\psi}$. Using the generalized Bloch theorem, the bulk equation indicates the ansatz $\psi_j=Ce^{i\theta j}\sin(qjd+\delta)$, with $q=k-\theta/d$. Substituting into the boundary conditions
\begin{subequations}
    \begin{align}
        -t e^{-i\theta}\psi_2-i\frac{\gamma}{2}\psi_1  &= E\psi_1,\\
        -t e^{i\theta}\psi_{N-1}-i\frac{\gamma}{2}\psi_N  &= E\psi_N
    \end{align}
\end{subequations}
and eliminating the parameters $C,\delta$ yields
\begin{equation}
    e^{2i(N+1)qd}=\left(\frac{2t-i\gamma e^{iqd}}{2t-i\gamma e^{-iqd}}\right)^2.
\end{equation}
Since we focus on the subradiant regime, we solve the above equation approximately near the band edge and obtain
\begin{equation}
    \begin{split}
        q_\xi&=\frac{\xi\pi}{Nd}\left[1-\frac{1}{N}+\frac{2\gamma^2}{N(4t^2+\gamma^2)}-i\frac{4\gamma t}{N(4t^2+\gamma^2)}\right]\\
        &\quad +\mathcal{O}\left(\frac{1}{N^3}\right),
    \end{split}
\end{equation}
with $\xi=1,2,\dots, \xi\ll N$. The corresponding decay rate $\Gamma_\xi=-2\text{Im}E_{k=\theta/d+q_\xi}=-2\alpha_2 \text{Im}q_\xi^2$ reads explicitly
\begin{equation}
    \Gamma_\xi=\frac{16\xi^2\pi^2(J_1^2+J_2^2+2J_1 J_2\cos\phi)\gamma}{[4(J_1^2+J_2^2+2J_1 J_2\cos\phi)+\gamma^2]N^3}.
\end{equation}
We show in Fig.~\hyperref[fgS:toy]{S11(c)} the $N^{-3}$ scaling of the minimal subradiant decay rate $\Gamma_\mathrm{min}=\Gamma_{\xi=1}$. More significantly, figure~\hyperref[fgS:toy]{S11(d)} demonstrates the decay rate as a function of the chirality parameter $\phi$. As $\phi$ increases and $|\alpha_2|$ delines, the subradiant decay rate can be suppressed correspondingly, which coincides with the mechanism we find regarding the BSs in chiral waveguide QED. Therefore, the results here indicates the universality of this subradiance mechanism.

\section{S-VI. Nanofiber interface \label{sec:Nanofiber}}
The details of the calculations regarding the nanofiber interface in the main text is given by this section. We clarify the modeling of the full atom-atom interaction, and then present the supplemental numerical results.

\subsection{A. Nanofiber-mediated interaction}
To depict the full interaction mediated by a nanofiber, we follow the theoretical treatment in Ref.~\cite{le2017nanofiber}. First, the propagation constant $k_0$ and the field distribution of the guided $\mathrm{HE}_{11}$ mode can be found in Appendix A of Ref.~\cite{le2017nanofiber}. See also Refs.~\cite{tong2004single,le2005nanofiber,solano2017optical}. For the infinite-range coupling arising from the guided mode, the single-atom decay rates are
\begin{subequations}
    \begin{align}
        \Gamma_{L}&=2\gamma_{jj}^{(\mathrm{g})-},\\
        \Gamma_{R}&=2\gamma_{jj}^{(\mathrm{g})+},
    \end{align}
\end{subequations}
with $\gamma_{jj}^{(\mathrm{g})\pm}$ same as Eq.~(21) of Ref.~\cite{le2017nanofiber}. The radiation-mode-mediated interaction is 
\begin{equation}
    H'=\sum\limits_{j,l=1}^N\left(\Omega_{jl}^{(\mathrm{r})}-i\frac{\gamma_{jl}^{(\mathrm{r})}}{2}\right) \sigma_j^\dagger \sigma_l,
\end{equation}
where $\Omega_{jl}^{(\mathrm{r})}$ and $\gamma_{jl}^{(\mathrm{r})}$ are given by Eqs.~(19) and~(20) in Ref.~\cite{le2017nanofiber}, respectively. Note that we also adopt the local-density-of-state approximation
\begin{equation}
    \Omega_{jl}^{(\mathrm{r})}\approx\frac{\sqrt{\gamma_{jj}^{(\mathrm{r})}\gamma_{ll}^{(\mathrm{r})}}}{\Gamma_\mathrm{f}}\Omega_{jl}^{(\mathrm{f})} 
\end{equation}
for simplicity, and we refer the reader to Ref.~\cite{svendsen2023modified} for a more accurate computation of $\Omega_{jl}^{(\mathrm{r})}$. Here, the free-space interaction is
 $\Omega_{jl}^{(\mathrm{f})}=-\mu_0\omega_\mathrm{f}^2\mathbf{d}_j^*\cdot\mathbf{G}_\mathrm{f}(\mathbf{r}_{jl},\omega_\mathrm{f})\cdot\mathbf{d}_l$, with~\cite{asenjo2017exponential}
\begin{equation}
    \begin{split}
        \mathbf{G}_\mathrm{f}(\mathbf{r},\omega_\mathrm{f})&=\frac{e^{ik_\mathrm{f}|\mathbf{r}|}}{4\pi k_\mathrm{f}^2 |\mathbf{r}|^3}\left[ (k_\mathrm{f}^2 |\mathbf{r}|^2 + i k_\mathrm{f} |\mathbf{r}| - 1)\mathbf{I} \right.\\
        &\left. \quad -(k_\mathrm{f}^2 |\mathbf{r}|^2 + 3i k_\mathrm{f} |\mathbf{r}| - 3)\frac{\mathbf{r}\mathbf{r}}{|\mathbf{r}|^2} \right]
    \end{split}
\end{equation}
and $k_\mathrm{f}=\omega_\mathrm{f}/c$. The single-atom natural linewidth satisfies $\Gamma_\mathrm{f}=\omega_\mathrm{f}^3|\mathbf{d}|^2/(3\pi\epsilon_0c^3)$. For an array of atoms with the identical radial coordinate $\rho_j=\rho_a$ and dipole momentum $\mathbf{d}_j=\mathbf{d}$, we denote $\Gamma'=\gamma_{jj}^{(\mathrm{r})}$, and define $\eta_0'=\Gamma_0/\Gamma'$ to characterize the cooperativity.

\begin{figure}[htbp]
    \includegraphics[scale=0.57]{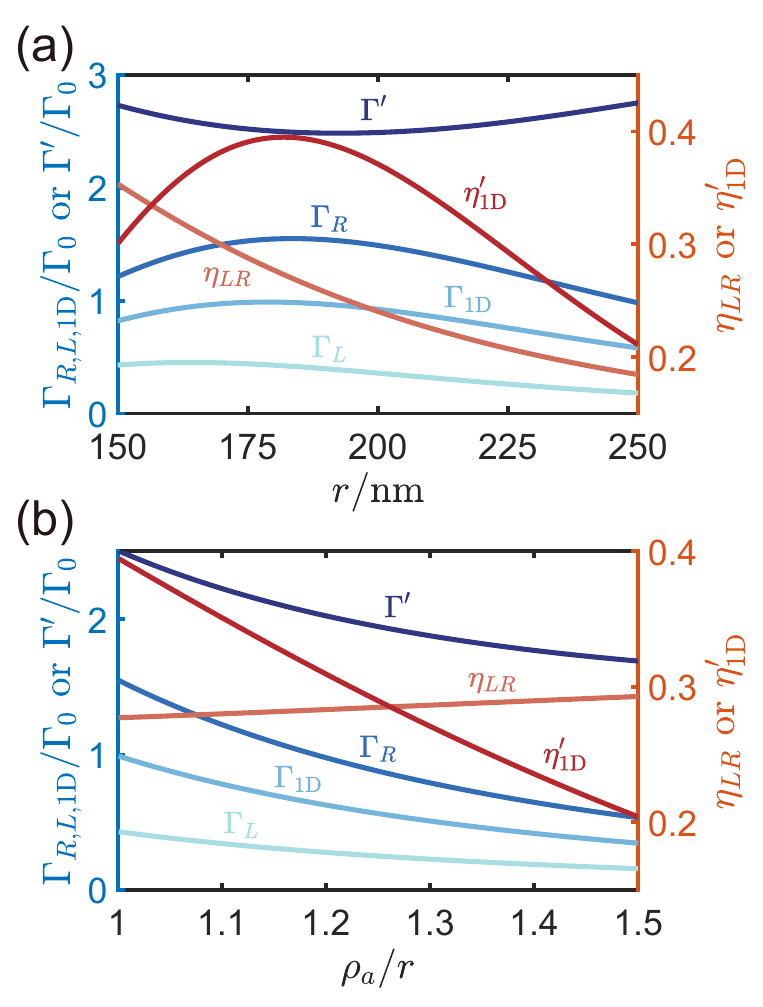}
    \centering
    \caption{Single-atom decay rates $\Gamma_L,\Gamma_R,\Gamma_0,\Gamma'$ and the parameters $\eta_{LR},\eta_0'$, as a function of the fiber radius (a) and the radial coordinate (b), respectively. Here, we fix $\rho_a=r$ in (a) and $r=180\,\text{nm}$ in (b).} 
    \label{fgS:ONFpurcell} 
\end{figure}

\subsection{B. Single-atom parameters}
Figure~\hyperref[fgS:ONFpurcell]{S12(a)} shows the decay rates $\Gamma_L,\Gamma_R,\Gamma_0,\Gamma'$ and the parameters $\eta_{LR},\eta_0'$ for a single $^{87}\mathrm{Rb}$ atom, as a function of the fiber radius $r$. Here, the atom is assumed to be located at the outer surface of the fiber, i.e., $\rho_a=r+0^+$, with other parameters same as the main text. Notably, the cooperativity $\eta_0'$ reaches a maximum at $r\approx 180\,\text{nm}$. Hence, we fix $r=180\,\text{nm}$ in the simulations below, as well as in the main text. In Fig.~\hyperref[fgS:ONFpurcell]{S12(b)} we show the dependence of these single-atom parameters on $\rho_a$. We find that as $\rho_a$ increases, $\eta_0'$ decreases dramatically, while $\eta_{LR}$ remains almost unchanged.

\subsection{C. Emergence of bound states from chirality}
\begin{figure}[htbp]
    \includegraphics[scale=0.55]{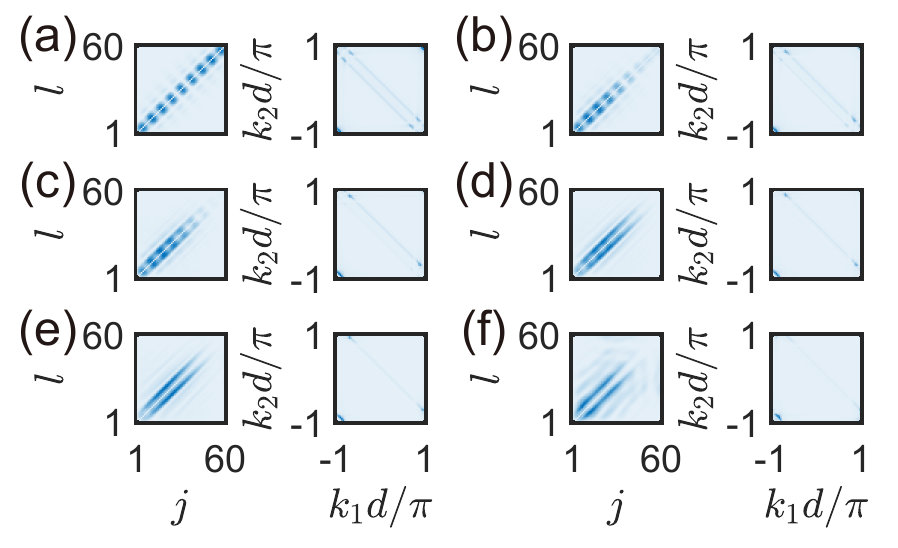}
    \centering
    \caption{Distributions of real space probability and momentum space probability amplitude for BS-like states in the nanofiber interface. The atomic dipole has $\mathbf{d}\propto \mathbf{e}_\rho, 4\mathbf{e}_\rho+i\mathbf{e}_x,3\mathbf{e}_\rho+i\mathbf{e}_x,2\mathbf{e}_\rho+i\mathbf{e}_x, (3/2)\mathbf{e}_\rho+i\mathbf{e}_x$ and $\mathbf{e}_\rho+i\mathbf{e}_x$ for (a-f), respectively. Other parameters are $N=60$, $r=180\,\text{nm}$, $\rho_a=r$ and $d=0.35\lambda_\mathrm{f}$.} 
    \label{fgS:ONFBS} 
\end{figure}

Here we show that the BS displayed in Fig.~\hyperref[fg:BSONF]{4(b)} of the main text originates from the chiral interaction. To this end, we consider the systems with $N=60$, $r=180\,\text{nm}$, $\rho_a=r$ and $d=0.35\lambda_\mathrm{f}$, tuning the chirality via varying the dipole orientation. For a non-chiral setup with $\mathbf{d}=\mathbf{e}_\rho$, we find a BS-like state with $9$ anitnodes in real space and two sets of iso-momentum lines in momentum space, as shown in Fig.~\hyperref[fgS:ONFBS]{S13(a)}. As chirality strengthens, with $\mathbf{d}\propto 4\mathbf{e}_\rho+i\mathbf{e}_x,3\mathbf{e}_\rho+i\mathbf{e}_x,2\mathbf{e}_\rho+i\mathbf{e}_x$ and $(3/2)\mathbf{e}_\rho+i\mathbf{e}_x $ in Figs.~\hyperref[fgS:ONFBS]{S13(b-e)}, respectively, the BS-like states exhibit increasingly well-defined features of standard bound photon pairs in chiral waveguide QED. For the case with circularly polarized dipoles $\mathbf{d}=(\mathbf{e}_\rho+i\mathbf{e}_x)/\sqrt{2}$, however, the state in Fig.~\hyperref[fgS:ONFBS]{S13(f)} show signs of mixing with the scattering resonances. Therefore, we focus on the state with $\mathbf{d}\propto(3/2)\mathbf{e}_\rho+i\mathbf{e}_x$ in the main text. Meanwhile, the BSs concentrate at the left edge, indicating their dissociation into the continuum. A more quantitative study may be based on Ref.~\cite{poddubny2026bound}.

\end{document}